\documentclass[conference]{IEEEtran}

\usepackage[T1]{fontenc}
\usepackage[utf8]{inputenc}
\usepackage{cite}
\usepackage{graphicx}
\usepackage{booktabs}
\usepackage{array}
\newcolumntype{L}[1]{>{\raggedright\arraybackslash}p{#1}}
\usepackage{amsmath}
\usepackage{xurl}
\newcommand{\um}{\ensuremath{\,\mu\mathrm{m}}}
\newcommand{\umsq}{\ensuremath{\,\mu\mathrm{m}^2}}

\newcommand{\code}[1]{{\ttfamily\def\_{\textunderscore\allowbreak}#1}}

\begin{document}

\title{Ankhdjet: An Open-Source Compiler for Mask-Programmed Ternary
Compute-in-ROM on an Open PDK}

\author{\IEEEauthorblockN{Mohnish Pai}
\IEEEauthorblockA{Independent Researcher\\
mohnishpai7@gmail.com}}

\maketitle

\begin{abstract}
Large-language-model inference is dominated by weight movement: every
generated token re-reads every weight. Ternary quantization (BitNet
b1.58) shrinks each weight to 1.58 bits with reported quality parity
at the 2B-parameter scale, small enough that hardwiring the weights
into a read-only mask becomes a plausible implementation, and a
commercial chip (Taalas HC1) has validated hardwired weights on an
advanced node behind closed tooling. This paper asks whether
model-specific silicon can be made reproducible with entirely open
infrastructure. We present Ankhdjet, an open-source compiler that lowers
a HuggingFace ternary checkpoint (BitNet b1.58 and its kin) to a
via-mask program of a fixed
compute-in-ROM macro on the open SKY130 process design kit, verified
end to end with open tools. We defend two claims: (1) the first
open-source weights-to-mask compute-in-ROM compiler on a fabricable
open PDK, taken through full open-toolchain signoff (KLayout DRC
zero, netgen LVS zero, clean timing) twice with two different weight
matrices through an identical flow in which only the mask differs;
and (2) the first compute-in-ROM macro submitted for fabrication on
an open PDK (TinyTapeout/ChipFoundry SKY130 shuttle ttsky26c, 2026
submission, silicon expected 2027). The compiler's read is
fully digital: full-swing bitline sampling into standard cells, so the
emission ports to any node by synthesis; an analog comparator readout
of the same machine is retained as a measured variant. We report the
comparison at every scale the artifacts allow: the digital readout
replaces the analog sense function at one eighth the area; the same
chip, on the same die and read contract, is signed off in both readout
styles through the identical flow, with the digital chip measuring
cleaner under the same rule engine; and a compiler-emitted digital
tile carrying the same mask program as the analog vehicle makes the
fabrication pair a controlled readout-style experiment awaiting
silicon. We additionally report the per-level adversarial
verification methodology that caught three DRC-invisible silent
shorts, a negative result on storing two ternary weights per
transistor at 130\,nm, and an analysis of which analog sense
techniques survive at which process nodes. Read energy is reported
from extracted-parasitic simulation (0.98--1.73\,pJ per sensed
weight across corners); no energy measurements are claimed, and the
fabricated vehicles are demonstrators.
\end{abstract}


\section{Introduction}
\label{sec:intro}

Autoregressive inference of a large language model re-reads the entire
weight set once per generated token. At edge batch sizes there is no
reuse to hide this behind: the working set is the model, and the
energy and latency of a token are set by how far the weights have to
travel. Software runtimes for ternary models push this wall back on
CPUs~\cite{wang2025bitnetcpp}, and FPGA accelerators specialize the
datapath around it~\cite{qiao2025tellme}, but as long as the weights
live in DRAM or even on-chip SRAM, every token pays to move them.

Ternary quantization changes what is worth building. BitNet b1.58
trains weights directly into $\{-1, 0, +1\}$ with per-tensor
scales~\cite{ma2024bitnet158}, and the 2B-parameter BitNet b1.58 2B4T
release reports quality comparable to full-precision peers of similar
size~\cite{ma2025bitnet2b4t}. At 1.58 bits per weight, two things
become true at once. First, a weight is small enough to be a
\emph{structural} choice: one transistor's drain either connects to a
positive bitline, to a negative one, or to neither. Second, the
multiply disappears: a ternary multiply-accumulate is an add, a
subtract, or a skip. Weights that never change can therefore be
manufactured rather than stored, in a mask-programmed read-only array
that is also the compute fabric.

There is now a commercial proof that hardwiring works at scale: the
Taalas HC1 hardwires Llama-3.1-8B on an N6-class node, with a
mask-programmable 1T ROM cell array described in the company's patent
filing, and a model swap that touches only two metal-layer
masks~\cite{taalas2025patent, taalas2026web, nextplatform2026taalas,
eetimes2026taalas}. The HC1 is closed on every axis that matters for
reproducibility: proprietary node, proprietary compiler, no published
artifacts. The academic side has the opposite gap: BitROM, the
closest published architecture, evaluates ternary compute-in-ROM at
65\,nm in layout only, with no compiler and no
silicon~\cite{zhang2026bitrom}.

This paper answers the question left between those two points: can an
open compiler, an open and fabricable PDK, and an open verification
toolchain make model-specific silicon a \emph{reproducible} artifact,
one that anyone can regenerate from a checkpoint, re-verify, and
submit to a foundry shuttle? We present Ankhdjet, an open-source compiler
that hardwires ternary weights into compute-in-ROM (CiROM) macros on
the SkyWater SKY130 open PDK~\cite{skywater}, carried through full
physical signoff with open tools only (Magic, KLayout, netgen,
ngspice, Yosys, OpenROAD, LibreLane)~\cite{magic, klayout, netgen,
ngspice, yosys, openroad, librelane}. One framing choice governs the
paper: ternary deletes the multiplier, so the machine is digital in
every variant (a ternary multiply-accumulate is an add, a subtract,
or a skip), and what the compiler actually emits is one of two
\emph{readout tiers} of the same architecture. The analog tier reads
a precharged bitline with a clocked comparator against a shared
reference; the digital tier samples the full-swing bitline into a
flop. Which tier wins is an empirical question this paper measures
where it can (Section~\ref{sec:tiers}) and otherwise leaves to the
fabricated instruments.

We defend exactly two novelty claims, scoped precisely in
Section~\ref{sec:related}:

\begin{enumerate}
\item \textbf{The first open-source weights-to-mask CiROM compiler on
a fabricable open PDK.} A HuggingFace ternary checkpoint (BitNet
b1.58 and its kin) lowers
to a ternary intermediate representation and then to a via-mask
program of a fixed, pre-verified macro. The flow reaches full
open-toolchain signoff: KLayout DRC zero, netgen LVS zero, clean
static timing. It has done so twice, with two different weight
matrices, through byte-identical configuration in which only the mask
differs; and it has done so in both readout styles: the same die and
read contract sign off with full-swing digital sampling and with the
analog comparator bands, the digital chip measuring cleaner under the
same rule engine.
\item \textbf{The first CiROM macro submitted for fabrication on an
open PDK.} The test vehicles \code{tt\_um\_darga\_cirom} (fully
digital readout, matrix-vector products computed on die) and
\code{tt\_um\_azara\_cirom} (the analog-readout characterization
instrument) target the TinyTapeout/ChipFoundry SKY130 shuttle
ttsky26c (2026 submission window), with silicon expected in 2027;
they carry the same mask program.
\end{enumerate}

Four further contributions support the claims. First, a verification
methodology (Section~\ref{sec:verification}): every hand-assembled
structure is flat-extracted and compared against its own generated
transistor-level schematic, a discipline that caught three silent,
DRC-legal shorts that chip-level signoff structurally cannot see.
Second, a measured readout-tier comparison
(Section~\ref{sec:tiers}): on the same signed-off implementation,
the digital sampler replaces the analog sense function at one eighth
the area, the comparators dominate the simulated read energy, and a
compiler-emitted digital tile carrying the same mask program as the
analog vehicle is hardened through the identical flow, and the full
chip is additionally signed off in both readout styles on the same
die; the analog
tier's remaining advantage is stated as an explicit hypothesis with
its test. Third, a negative result (Section~\ref{sec:biroma}):
BitROM's two-ternary-weights-per-transistor cell does not port
economically to 130\,nm; we give a topological disproof for one
variant and a measured rail-droop argument for the other, and we
bound the node window in which the technique pays. Fourth, a
node-handoff analysis (Section~\ref{sec:scaling}): the analog
techniques in this design (offset oversizing, multi-level sensing,
charge-domain binary read) each expire at a different process node,
and we map which crossings change the architecture rather than merely
the layout.

Throughout, numbers are reported with their provenance. Silicon-flow
numbers come from signoff runs of record; simulation numbers name the
corner and the extraction level; estimator numbers are labeled as
calibrated model estimates. Energy and power appear only as
extracted-parasitic simulation and labeled vectorless reports
(Section~\ref{sec:energy}); nothing is measured, and no efficiency
victory over measured silicon is claimed.

\section{Related Work}
\label{sec:related}

Ankhdjet sits at the intersection of three research threads, none of
which individually contains it. Table~\ref{tab:positioning} summarizes
the positioning; the text below states explicitly what each prior work
established and what Ankhdjet does and does not claim beyond it.

\begin{table*}[t]
\caption{Positioning against prior work. ``Weights bound'' states when
weight values become fixed. Silicon status is as published. Ankhdjet's
claims are exactly the last row's two boldface items; everything else
in the table is prior art we build on or differentiate from.}
\label{tab:positioning}
\centering
\footnotesize
\setlength{\tabcolsep}{4.5pt}
\begin{tabular}{@{}llllll@{}}
\toprule
Work & Weight storage & Weights bound & Node / PDK & Compiler released & Silicon status \\
\midrule
BitROM~\cite{zhang2026bitrom} & 1T ROM, 2 wt/cell & mask & 65\,nm (commercial) & no & none (layout eval.) \\
TOM~\cite{guan2026tom} & std-cell logic + ROM & synthesis & 7\,nm (commercial) & no & none (simulation) \\
DCiROM~\cite{yu2025dcirom} & digital CiROM & mask & commercial & no & none (macro eval.) \\
AutoDCIM~\cite{chen2023autodcim} & SRAM CIM & runtime & TSMC 40\,nm & yes (closed) & none reported \\
SynDCIM~\cite{shao2025syndcim} & SRAM CIM & runtime & 40\,nm (commercial) & yes & test chip (commercial) \\
OpenC2~\cite{dong2025openc2} & SRAM CIM & runtime & FreePDK45 (predictive) & yes (open) & none (predictive PDK) \\
OpenACM~\cite{zhou2026openacm} & SRAM approx. CIM & runtime & commercial/predictive & yes (open) & none reported \\
OpenSpike~\cite{modaresi2023openspike} & SRAM (OpenRAM), SNN & runtime & SKY130 (open) & RTL + flow open & taped out (MPW) \\
rejunity~\cite{rejunity2024bitnet} & digital registers & runtime & SKY130 (open) & RTL open & fabricated (TT06 2024) \\
T3~\cite{t3tt09} & digital registers & runtime & SKY130 (open) & RTL open & fabricated (TT09 2024) \\
Slim-Llama~\cite{kim2025slimllama} & on-chip digital memory & runtime & 28\,nm (commercial) & no & fabricated \\
Taalas HC1~\cite{taalas2025patent, taalas2026web} & 1T mask ROM & mask & N6 (commercial) & no (proprietary) & fabricated, commercial \\
\midrule
\textbf{Ankhdjet (this work)} & 1T NOR mask ROM & \textbf{via-1/jog mask} & \textbf{SKY130 (open, fabricable)} & \textbf{yes (open)} & signoff $\times$2; submitted 2026 \\
\bottomrule
\end{tabular}
\end{table*}

\subsection{ROM-Based Compute Architectures}

Mask ROM as a dense, manufacturable memory has decades of production
precedent; the relevant modern anchors are a 16\,nm FinFET 1T ROM
macro with low-swing clocked sensing~\cite{verma2015rom} and a 40\,nm
16\,Mb contact-programmed mask ROM~\cite{ye2016maskrom}. Computing in
ROM is newer. DCiROM builds fully digital compute-in-ROM macros for
task-level DNN inference~\cite{yu2025dcirom}. BitROM is the closest
prior work to this paper: it defines a ternary CiROM architecture
(one-transistor cells storing two ternary weights each via the BiROMA
encoding, tri-state multi-level accumulation, a shared adder tree)
targeted at billion-parameter BitNet inference~\cite{zhang2026bitrom},
with architecture code published~\cite{zhang2025bitromgithub}. BitROM
is evaluated at 65\,nm in layout and simulation; it does not include a
weights-to-mask compiler and has no silicon. TOM takes a different
route to the same goal: ternary weights become synthesized
standard-cell logic feeding separate MAC units, evaluated in 7\,nm
simulation~\cite{guan2026tom}. Ankhdjet does not claim the ternary
weights-in-ROM idea: BitROM published it first, and our array
architecture deliberately stays close to BitROM where BitROM has done
the analysis. What no prior work in this thread has is a compiler
from a public checkpoint to a mask program, an open fabricable PDK
implementation, signoff with open tools, or a fabrication submission.

\subsection{Compute-in-Memory Compilers}

A recent line of work automates CIM macro generation. AutoDCIM
compiles digital SRAM-CIM macros~\cite{chen2023autodcim}; SynDCIM adds
performance-aware subcircuit synthesis and validates with a 40\,nm
test chip~\cite{shao2025syndcim}; OpenC2 is an open-source end-to-end
DCIM compiler framework demonstrated on the predictive FreePDK45
kit~\cite{dong2025openc2}; OpenACM extends open CIM compilation to
approximate computing~\cite{zhou2026openacm}. These establish that CIM
macro compilation can be automated, and OpenC2 and OpenACM establish
that it can be open. All four, however, compile \emph{SRAM-based,
runtime-programmable} macros, on commercial or predictive
(non-fabricable) processes, and none compiles the \emph{weights
themselves} into the layout. Ankhdjet is not the first CIM compiler and
not the first open one; it is the first whose output is a
mask-programmed model (the weight values are the artifact being
compiled) on a PDK that a foundry actually accepts.

\subsection{Accelerator Silicon on Open PDKs, and Ternary Silicon}

OpenSpike taped out a spiking neural network accelerator on SKY130
with OpenRAM memories and a fully open flow~\cite{modaresi2023openspike,
openram}, establishing open-EDA NN-accelerator tapeout before this
work. On the ternary side, a TinyTapeout TT06 project implemented a
BitNet-style 1.58-bit matrix multiplier as digital
logic~\cite{rejunity2024bitnet}, and T3 followed on
TT09~\cite{t3tt09}; both hold weights in registers at runtime. At the
commercial end, Slim-Llama demonstrated a fabricated binary/ternary
LLM processor at ISSCC 2025~\cite{kim2025slimllama}. TinyTapeout has
also carried conventional (non-computing) mask-ROM
experiments on SKY130~\cite{ttromexp}. Ankhdjet claims none of these
firsts: not the first NN accelerator on an open PDK, not the first
ternary arithmetic on open-PDK silicon, not the first ternary-LLM
silicon. Its second claim is precise: no compute-in-ROM macro (weights
manufactured into the mask, sensed and accumulated in place) has
previously been submitted for fabrication on an open PDK.

\subsection{Taalas}

Taalas is the commercial existence proof for the whole premise. The
HC1 hardwires an 8B-parameter model into an N6 die; public statements
describe roughly 53\,B transistors serving 8\,B parameters, a
mask-programmable 1T cell with via-selectable bitline connection, and
model swaps confined to two metal masks~\cite{taalas2025patent,
taalas2026web, nextplatform2026taalas, eetimes2026taalas}. In August 2026 AMD announced a definitive
agreement to acquire Taalas, moving hardwired-model silicon onto a
major vendor's accelerator roadmap~\cite{amd2026taalas}. The filing
deliberately stops at the sense-amplifier output; the post-sense
compute fabric is undisclosed. The filing is precedent that the
structural family (mask-programmed 1T ROM, via-selectable bitline
connection) is production-viable; Ankhdjet belongs to that family but
differs where the filing claims: one weight per cell with no shared
bitline connections (the filing's claimed density mechanism) and an
opposite sense scheme (precharge-high against VDD/2 versus the
filing's comparison against a precharge-low value). Where the filing
is silent, Ankhdjet adopts BitROM's published choices, and it is in
every other respect the opposite experiment: minimum viable node
instead of N6, and every artifact open instead of none.

The deeper relationship is that the two designs place every
operation of a decode step identically: parameters in mask ROM, read
locally once per token; attention scores, softmax, and normalizations
in weight-free digital units dimensioned for the model; the KV cache
in on-die writable memory, the one block no amount of area converts
into masks. The fork is confined to the sense boundary, to what a
sensed bitline becomes. Taalas restores it to a digital bit at the
sense amplifier and multiplies in logic; Ankhdjet lets the ternary
symbol be decided by the sense itself, a comparator against VREF
classifying which line discharged. Both are ADC-free, dodging the
converter from opposite sides. Section~\ref{sec:scaling} shows one
direction of the choice between them: below roughly 28\,nm the analog
tier trades device oversizing for auto-zeroing and its area advantage
erodes, tilting the balance digital; above it both sides work, and the
choice is a
workload-dependent energy tradeoff this paper does not resolve.

\section{The Compiler and the Mask-Only Contract}
\label{sec:compiler}

\begin{figure*}[t]
\centering
\includegraphics[width=0.92\textwidth]{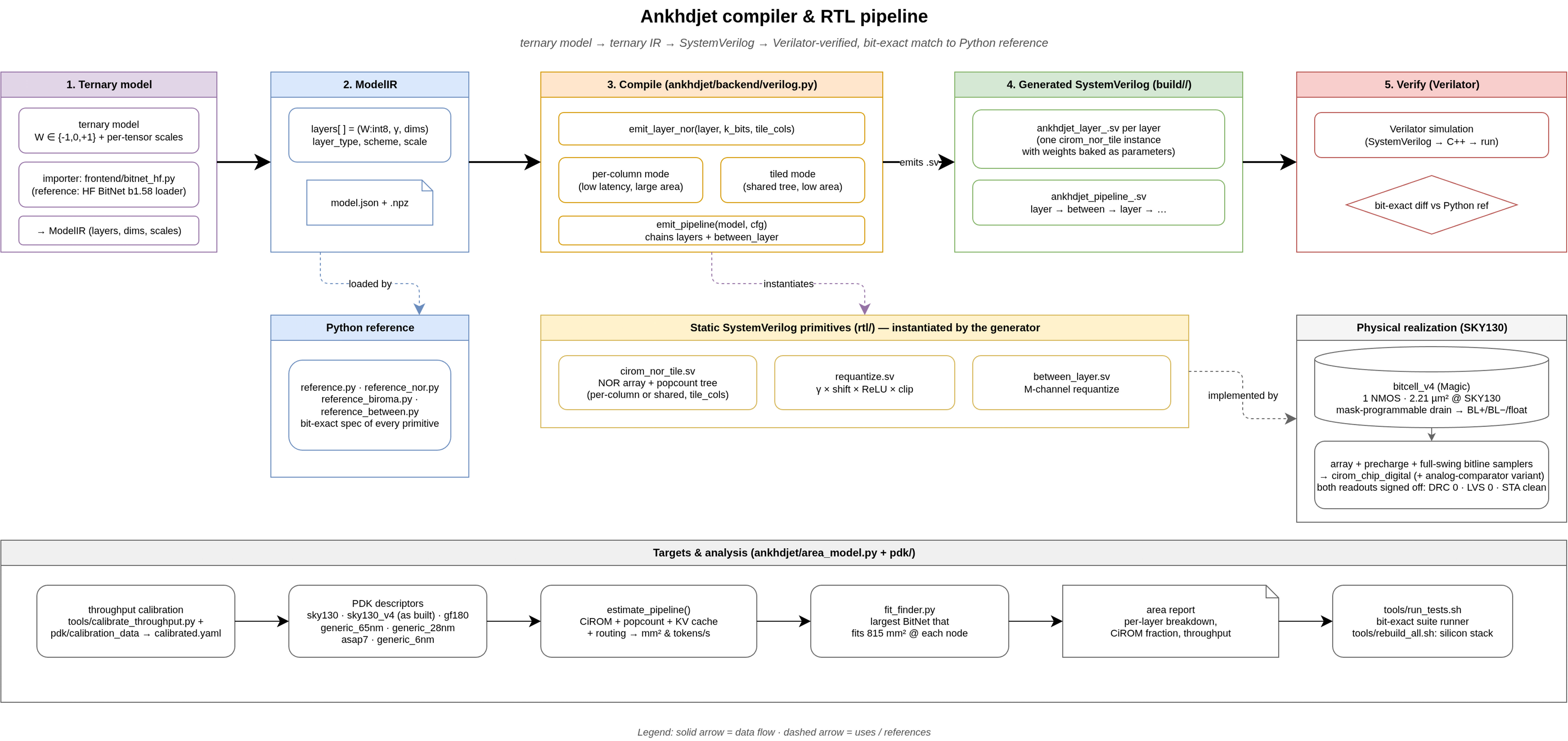}
\caption{The Ankhdjet flow: checkpoint to ternary IR, per-layer
wrapper RTL plus a mask program for a fixed CiROM macro, then the
open physical flow to signoff. Weight values enter the layout only
at the mask-programming step.}
\label{fig:pipeline}
\end{figure*}

The compiler is the installable half of the toolchain: a
pip-installable Python package whose input is a checkpoint and whose
output is a design bundle, two re-expressions of the same model plus
the collateral to place and check them (Figure~\ref{fig:pipeline}). The bundle holds the
per-layer mask programs (the fab handoff), a self-contained RTL view
of the same design, and per-shape physical abstracts of the macro
the mask programs assume, every part covered by content digests. The
physical machinery that turns a mask program into hardened GDS (the
array generator of Section~\ref{sec:macro}, the LibreLane flows of
Section~\ref{sec:results}) stays with the repository, related to the
package the way a linker is related to a compiler: the package
proves what should be built and hands over everything needed to
build it and to check the build.

\subsection{Frontend: Checkpoint to Ternary IR}

The frontend ingests any HuggingFace checkpoint whose matmul weights
are ternary; BitNet b1.58 is the headline family (the reference
target is the Microsoft 2B-4T release~\cite{hf2025bitnetmodel}), not
a requirement. Two
entry points exist: a configuration-only path that builds the model IR
from \texttt{config.json} (sufficient for area and throughput
reporting), and a full-weight path that decodes the safetensors
payload, detecting the storage format per tensor. Three formats
decode: the packed two-bit-lane encoding of the transformers BitNet
integration (four ternary values per byte, per-tensor scale
alongside; the storage of the BitNet, Falcon-E, and ternarized-Llama
releases), ternary-valued float storage whose sign pattern and
embedded scale are recovered directly (the TriLM-class unpacked
releases), and, on explicit request only, QAT master weights through
the b1.58 absmean transform. A tensor that is none of these is
refused as non-ternary rather than silently quantized, a posture
with a measured justification: applying absmean to the flagship's
released bf16 master weights reproduces only 98.5\% of the packed
release's ternary values (32\,M of 2{,}084\,M positions flip on
rounding boundaries), so coercion yields a near twin of the model,
not the model. The IR records each
matmul-resident layer (attention projections, MLP projections, the
final vocabulary projection) as an $N \times M$ ternary tensor plus
quantization metadata. Attention itself (softmax over $QK^{T}$) is not
a linear layer and is outside the scope of the ROM arrays.

Two guardrails govern ingestion. The bf16-tied output projection is
off-fabric by contract: it is an embedding-tied table rather than
ternary silicon, so the frontend records it in the model manifest as
an off-fabric parameter count instead of silently dropping it or,
worse, silently including it. And a layer whose weights cannot be
loaded keeps its correct dimensions behind a loudly flagged
placeholder, so area estimation keeps working while emission and
verification refuse the layer.

\subsection{Backend: Mask Programs}

The backend tiles each layer's $N \times M$ ternary matrix into
macro-shaped chunks, zero-padding ragged edges; a padded position is
not a modeling convenience but a physical zero weight, a drain
connected to neither bitline. Each chunk becomes a plain-text matrix
of \texttt{+/-/0} characters, one per cell. That deliberately
primitive format is the entire interface to physical reality: the
array generator consumes it and places each cell's via and jog
choice, which is the entire per-model mask difference. A per-layer
manifest records grid geometry, padding, and a content digest per
chunk, and a model manifest records the totals; the digests make
drift between what was compiled and what sits on disk detectable
long after the fact.

\subsection{Backend: The RTL View}

The second re-expression is SystemVerilog, and one principle governs
it: emitters generate \emph{structure, never logic}. All behavior
lives in a hand-written, pre-verified RTL library (the ternary tile
primitive, the between-layer requantize, the grid readout
controller); the compiler emits instances and parameters over that
library. Per layer that means a wrap of the tile with the via
program baked in as two packed bit vectors (\code{HAS\_VIA\_POS},
\code{HAS\_VIA\_NEG}), a multi-layer pipeline chained through the
requantize stages with a valid/start handshake, and a grid top that
sweeps the tiled macro grid row-chunk by row-chunk while
accumulating every output column in parallel. A compiler bug
therefore cannot manufacture novel logic; it can only
mis-parameterize modules the bit-exact suite has already pinned,
which is exactly the failure class the reference chain below is
built to catch. Every name that crosses from the checkpoint's
namespace into RTL is legalized and collision-checked at the
emission boundary.

The bundle is self-contained by construction: the library sources
are copied beside the generated tops, together with behavioral
\texttt{.memh} views of the same mask programs, a parameterized
smoke bench, and a file list, so the emitted design elaborates and
simulates standalone on a machine that has never seen the
repository. A digest manifest covers the whole bundle.

\subsection{Physical Abstracts and PDK Packs}

For floorplanning the bundle carries per-shape macro abstracts: a
LEF footprint, a Liberty timing and power view, and a blackbox
Verilog module, computed for the exact grid shape the model compiled
to. They are generated from a \emph{PDK pack}: a data-only directory
whose manifest declares capability tiers (estimator descriptors;
abstract-generation constants with a characterized anchor macro;
physical collateral), each tier optional and declared independently.
A process with calibrated estimators but no hardened macro therefore
ships as an estimate-only pack, and a request against a tier a pack
does not declare is refused with a diagnosis rather than served from
an invented template. Conformance is checked per declared tier and
measured rather than asserted: an abstracts-tier pack's constants
must regenerate its anchor macro's
extracted LEF rectangle for rectangle, and Liberty capacitances
scale from anchor-extracted values. The abstracts are template-grade
collateral for early floorplanning; the hardened macro's own
extracted views remain the signoff collateral. The reference
distribution carries three packs: SKY130 with all three tiers
anchored to the signed-off macro, and estimate-only GF180 and ASAP7
packs whose descriptors carry the measured characterization and
synthesis anchors.

Packs resolve from an environment path, then from installed packages
exposing a pack entry point, then from the copy bundled with the
compiler. A commercial PDK therefore plugs in as a private pack from
a private index, carrying its own anchors and physical collateral,
without any change to the open compiler; a restricted flag on such a
pack propagates into every manifest emitted from it. When a pack
carries physical collateral, a hardened binding mode re-emits the
same grid tops over per-chunk hard-macro instances with identical
interfaces, plus per-layer blackboxes and a manifest mapping every
chunk to its macro and mask program: the netlist a foundry
place-and-route flow consumes directly.

\subsection{The Invariant: Weights Live Only in the Mask}

\begin{figure}[t]
\centering
\includegraphics[width=0.92\columnwidth]{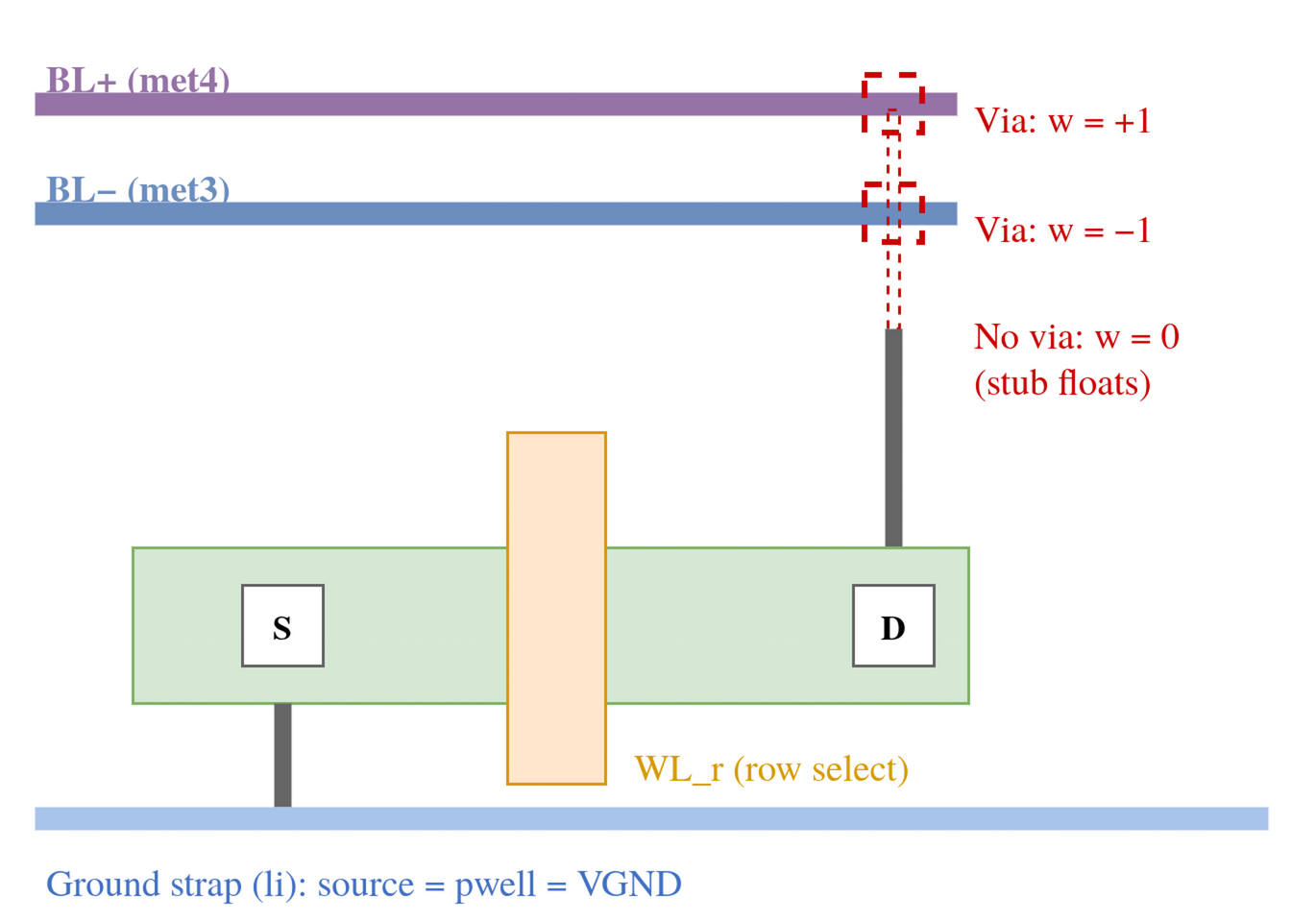}
\caption{The weight is a single drain-via choice: BL+ for $+1$,
BL$-$ for $-1$, none for $0$. Every other terminal role, and every
other mask, is identical across models.}
\label{fig:bitcell-via}
\end{figure}

The central contract is that a weight matrix affects exactly one
thing: the via-1/metal-jog finishing geometry of the array
(Figure~\ref{fig:bitcell-via}). Cell
placement, bitline and wordline geometry, precharge devices, sense
macros, pin positions, the LEF abstract, the Liberty timing view, the
floorplan, and the flow configuration are all invariant across weight
matrices of the same shape. One verified abstract therefore serves
every variant of a shape, and recompiling a model is a mask respin,
not a re-implementation. Via-programmable ROM practice has worked this way for decades, and
the Taalas two-mask model swap is the commercial
precedent~\cite{taalas2025patent}.
Section~\ref{sec:results} shows the invariant exercised for real: two
signoffs, two matrices, one flow.

\subsection{Bit-Exact Reference Chain}

Numerical correctness is anchored by a chain of four models, each
compared exactly (integer equality, no tolerance) against the next:

\begin{enumerate}
\item \textbf{PyTorch}: \texttt{F.linear} on the unpacked ternary
tensors, the frame of reference users care about.
\item \textbf{Pure-math oracle}: a NumPy bit-serial ternary dot
product with no array structure; validates the unpacking and the
bit-serial decomposition ($\mathrm{acc} = \sum_k \mathrm{partial}_k
\ll k$) against PyTorch.
\item \textbf{Hardware-order oracle}: a row-sequential, one-hot
wordline, column-tiled model that mirrors the physical readout order
(precharge, single-row evaluate, per-column hit, per-bit-slice
accumulate). An earlier parallel-wordline abstraction was discarded
when it proved impossible to implement with 1T cells; the row-sequential model
is the ground truth the RTL is held to.
\item \textbf{RTL}: Verilator simulation of the emitted SystemVerilog,
checked cycle-by-cycle against the hardware-order oracle, including on
weight slices taken from the real 2B4T checkpoint.
\end{enumerate}

The chain is closed at the artifact level as well. The same weights
leave the compiler by three independent routes (parameter bit
vectors in the RTL, \texttt{.memh} views, mask programs), and the
routes are checked against one another. An audit command reassembles
every layer's mask programs from what is actually on disk and
compares them bit for bit against a fresh unpacking of the
checkpoint, failing on any content mismatch and on any nonzero value
in a padded position; the token-parity twin of
Section~\ref{sec:asap7} extends the same discipline from tensors to
generated text.

The same expectations flow downward into the physical world: the
array generator emits, for every weight matrix, both the mask program
and a transistor-level schematic whose netlist the extracted layout
must match (Section~\ref{sec:verification}), and the functional bench
reads its expected values from views generated from the same matrix.

\section{The SKY130 CiROM Macro}
\label{sec:macro}

The hardened macro family implements a 1T NOR CiROM array with
clocked precharge. The NOR topology is a deliberate choice for a
matrix-vector primitive: each cell sits in parallel between its
bitline and ground, so a single asserted wordline discharges every
selected column's bitline independently in one step, delivering the
full-parallel random-row read the per-token matvec depends on. The
denser NAND alternative wires cells into series strings read through
the entire stack, trading that read speed for density: a smaller-node
density lever, not the primitive used here. Everything up to the bitline (cell, mask
programming, source straps, precharge) is shared by both readout
tiers and is weight-invariant; the tiers differ only at the last
step, how a bitline's state becomes a bit. The readout of record is
digital: wait for the full swing and sample the bitline into a flop
through an input buffer, entirely standard cells, with no analog
signoff surface. The analog variant replaces the samplers with a
per-bitline clocked comparator against a shared reference, and
survives as the measured counter-hypothesis
(Section~\ref{sec:tiers}). The section is organized along that
split: the shared structure first (bitcell, precharge), then the
digital readout, then the analog variant's sense stack, which
carries all of the macro's analog signoff machinery.

This section describes the silicon-verified configuration: a $64 \times 32$ array
(2{,}048 ternary weights), 64 rows per bitline, read in a
five-state precharge/evaluate/strobe/hold cycle at a 25\,ns clock
(125\,ns per read). The shape is chosen
as the smallest configuration that exercises every architectural
element, not for density: 64 rows is a full 6-bit one-hot decode at
the bitline depth the extracted-netlist read validation and the
Monte Carlo campaign cover end to end; 32 columns places both
polarities' 64 bitlines behind exactly four of the hardened
16-comparator bands (two in the analog vehicle's scaled subset); and
the identical macro drops unchanged into the chip die and the
$1{\times}2$ shuttle tile, so every vehicle shares one
silicon-verified array. Production geometry tiles deeper, at 128--256
rows per bitline (Section~\ref{sec:scaling}).

\subsection{Bitcell}

\begin{figure}[t]
\centering
\includegraphics[width=0.94\linewidth]{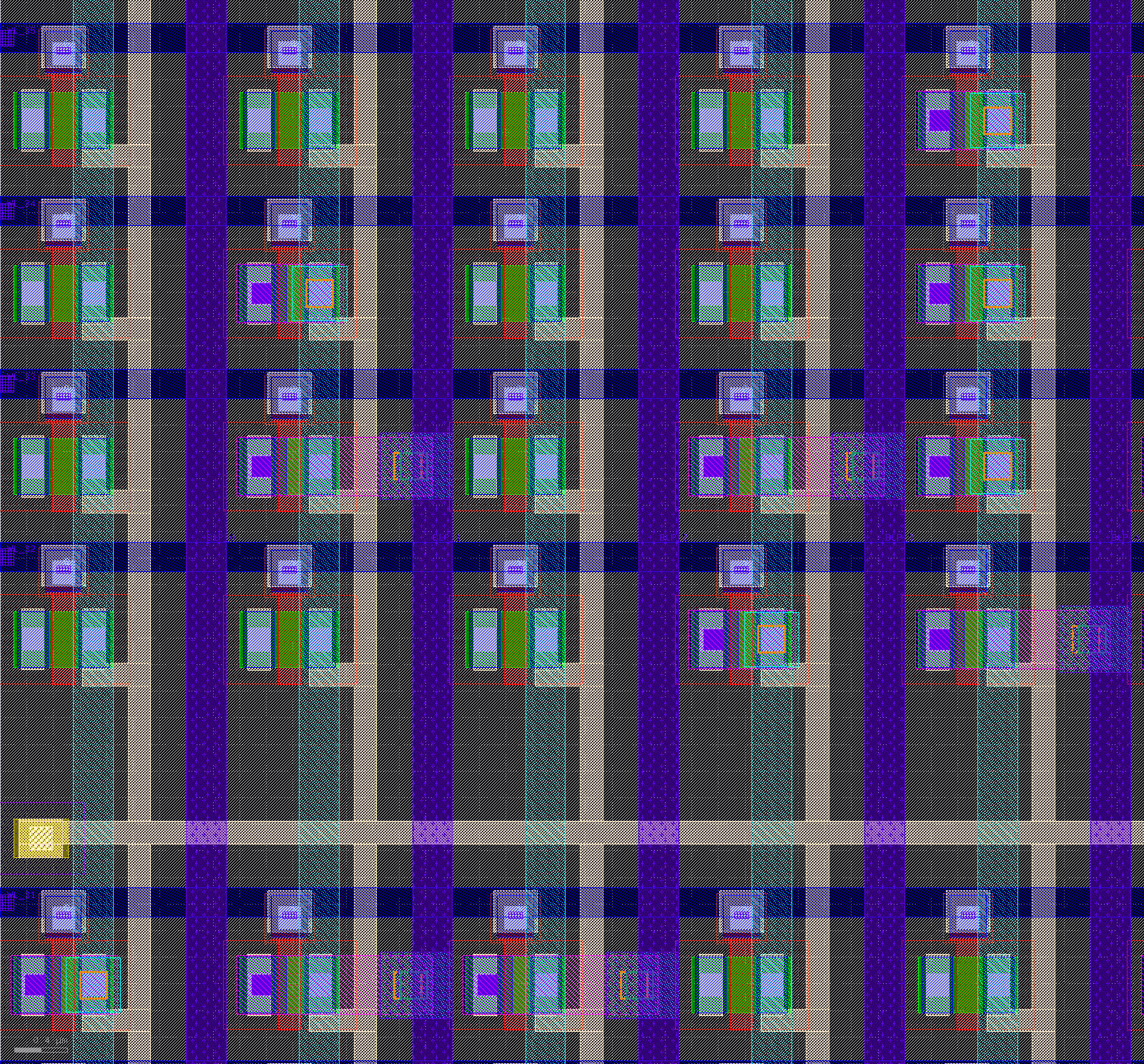}
\caption{Mask programming in the 1T NOR array. Per cell, one choice
from \{BL$+$, BL$-$, none\} is written as via/metal-jog finishing
geometry; the weight matrix is otherwise invisible to the layout. The
abutment rule for adjacent jogs (Section~\ref{sec:verification}) came
out of a caught silent short.}
\label{fig:maskjogs}
\end{figure}

Each cell is a single NMOS (\code{sky130\_fd\_pr\_\_nfet\_01v8},
$W = 0.42$, $L = 0.17$) at a 1.70\um{} $\times$ 1.30\um{} pitch:
2.21\umsq{} per ternary weight as built. The gate is the row wordline;
the source joins a per-column local-interconnect strap network tied
through tap rows to the substrate, so sources, p-well, and ground
extract as one net; the drain is the programmed terminal. A weight of
$+1$ routes the drain through a metal jog to the column's BL$+$
strip, $-1$ to the BL$-$ strip, and $0$ leaves the drain stub
floating: mask-absent, modeled as a dangling net in the generated
schematic and verified as such by extraction
(Fig.~\ref{fig:maskjogs}). Body taps occupy a separate tap row every
eight rows.

Measured on PDK models at the slow corner (\texttt{ss}, 100$^{\circ}$C,
1.62\,V), a selected cell discharges its bitline to 0.9\,V in 0.54\,ns
and to 0.3\,V in 1.07\,ns, against an 8\,ns evaluate state: a margin
of better than $4\times$ on every timing constraint of the read.
Extracted from the full 2{,}048-cell macro, wire-only bitline
capacitance is 27--37\,fF; the dominant coupling is the in-column
BL$+$/BL$-$ pair at 7.27\,fF (roughly 25\% of the bitline
capacitance), and the grounded source straps shield every other
pairing below 0.8\,fF.

\subsection{Clocked Precharge, No Keeper}

Both bitlines of every column carry one PMOS pull-up ($W/L =
1.0/0.15$), gated by an active-low precharge signal from the read
FSM; bitlines park at the rail whenever the array is idle. The
decision to clock the precharge rather than use a static pseudo-NMOS
pull-up came from a measurement, not a preference: a pull-up weak
enough for tolerable static power cannot re-charge the bitline within
the 25\,ns cycle at the slow corner (21.7--43.3\,ns to 90\% of the
rail for $L{=}1$/$L{=}2$ devices), while the clocked device recovers
the line in 1.63\,ns worst-case. There is no keeper: worst measured
leakage on a dynamically held line (64 off cells, fast corner,
100$^{\circ}$C) droops it 0.88\,mV over a full cycle, three orders of
magnitude inside the sense margin, and the read style (single one-hot
row, re-precharged every cycle) is inherently leakage-tolerant. The
same conclusions are embodied in OpenRAM's ROM precharge (clock-gated
PMOS)~\cite{openram} and BitROM's precharge-equalize
stage~\cite{zhang2026bitrom}.

\subsection{Digital Readout of Record: Full-Swing Sampling}
\label{sec:digital-readout}

The digital tier's sense is one input buffer and one capture flop
per bitline: the read FSM's strobe clocks the flop, and a discharged
bitline has crossed the buffer's switching threshold well inside the
evaluate state (the bitcell's better-than-$4\times$ slow-corner
margin above is measured to that full-swing crossing). Ternary is
what makes this sufficient: with a single one-hot row driving the line,
the read is a digital 1-of-2 event per bitline, ($0,0$) is a legal
codeword by construction, and there is no millivolt-scale decision
anywhere, hence no reference, no mismatch signoff, and no Monte
Carlo obligation. The samplers and the downstream accumulation
synthesize with the periphery against the standard-cell library, so
the custom silicon ends at the array-plus-precharge boundary and the
macro contract owes the periphery nothing but full-swing bitlines
and the strobe timing. The area consequence of deleting the
mismatch tax is measured in Section~\ref{sec:tiers}: roughly
68\umsq{} per weight column of sampling logic against 540\umsq{}
for the comparator bands it replaces.

\subsection{Analog Variant: Why Two Single-Ended Comparators per
Column}
\label{sec:sense}

The sense design is built around a correctness result rather than a
performance one. A ternary column read has three outcomes: BL$+$
discharged ($+1$), BL$-$ discharged ($-1$), neither ($0$). A
differential comparator across (BL$+$, BL$-$) regenerates to
complementary outputs: it can emit $(1,0)$ or $(0,1)$ but never
$(0,0)$, so the zero state, which presents both bitlines high and a
near-zero differential, resolves by the latch's own offset to a
random but per-column-coherent $\pm 1$. BitNet matrices are
zero-heavy (the committed test matrix is 50\% zeros), so a
differential front end silently corrupts a large fraction of every
dot product. No sizing fixes a representation gap. The implemented
sense is therefore two independent single-ended clocked comparators
per column, each answering ``did this bitline discharge?'' against a
shared reference $\mathrm{VREF} \approx \mathrm{VDD}/2$, with $(0,0)$
a legal codeword. The same structural conclusion is visible in prior
art: BitROM senses single-ended against fixed VDD-fraction
references~\cite{zhang2026bitrom}, and TOM avoids analog sensing
entirely~\cite{guan2026tom}.

The comparator is a StrongARM-derived clocked
latch~\cite{razavi2015strongarm} with one input tied to VREF. At
130\,nm the mismatch tax dominates its design. Monte Carlo against
the PDK mismatch corner puts the bare minimum-size input pair
($W/L = 2/0.3$) at $\sigma_{\mathrm{offset}} \approx 130$\,mV at the
slow corner, against a $\pm 50$\,mV budget derived from the reference
accuracy stack-up. Pelgrom-style area scaling closes the gap: the
implemented input pair is $W = 8$\um{}, $L = 2$\um{} (16\umsq{} of
gate area per device), and the implemented comparator passes 250/250
mismatch Monte Carlo trials at realistic bitline levels for both the
discharged and the weight-zero case. The oversizing is not free, in
three measured ways. First, each input presents about 133\,fF of gate
capacitance, so the comparator, not the wire, dominates the bitline
load. Second, 64 comparators strobing synchronously kick roughly
3.5\,pC back into the shared VREF node, which sets a local decoupling
requirement (Section~\ref{sec:sense-vref}). Third, the latch races
its two output nodes, and with the internal complement node nearly
unloaded, decisions flip when the routed output load reaches about
150\,fF (measured on the extracted macro at the slow corner); the
Liberty view therefore carries a 60\,fF ceiling on the HIT pins and
the flow keeps a buffer placed directly under each one.

\subsubsection{Reference}
\label{sec:sense-vref}

The reference is ratiometric by design: bitlines precharge to VDD, so
the threshold should track VDD across supply corners, which a
VDD/2 divider does and an absolute reference would not (BitROM
likewise specifies its comparator references as VDD
fractions~\cite{zhang2026bitrom}). Both fabricated vehicles take VREF
from an external pin. An on-chip divider (two 180\,k$\Omega$
high-resistivity poly legs, interdigitated) with $\geq$150\,pF of
split MiM decoupling was designed and validated on the extracted
netlist with all 64 comparators strobing synchronously (dips of
18--22\,mV across corners, verdicts correct, against the
$\pm 50$\,mV budget), but is not instantiated in the current
assemblies; the external pin doubles as a bring-up instrument, since
sweeping it ramp-converts every column's bitline voltage through the
comparators.

\subsection{The Analog Sense Band}

\begin{figure}[t]
\centering
\includegraphics[width=0.72\linewidth]{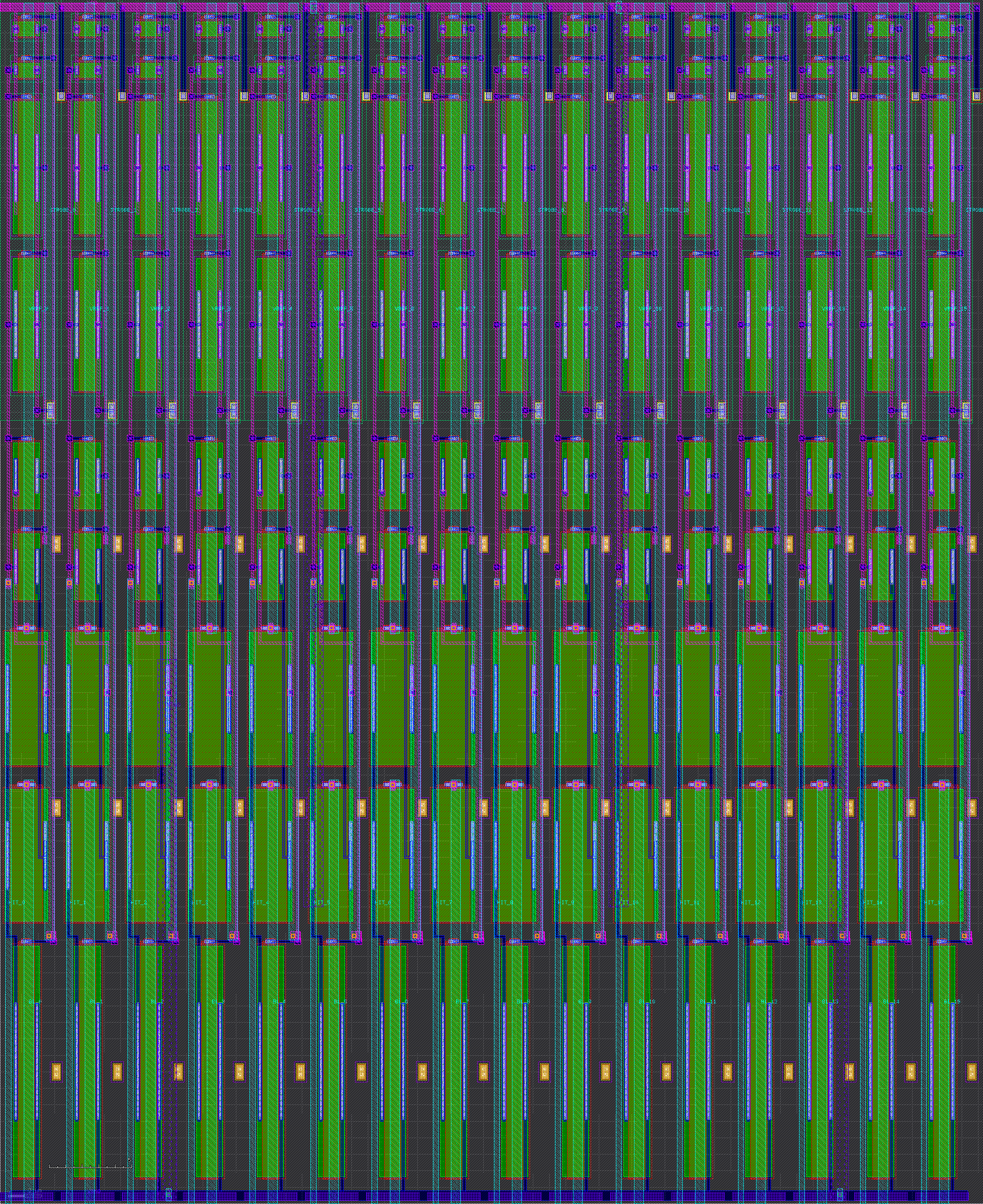}
\caption{The 16-comparator sense band macro (59.3\um{} $\times$
73.0\um{}). Metal-3 risers carry buried bitline, VREF, strobe, and
HIT pins to the boundary; the band flat-extracts and must match its
generated schematic before any build that contains it may publish.}
\label{fig:band}
\end{figure}

Sixteen comparators tile into one flat-verified band macro of
59.3\um{} $\times$ 73.0\um{}, about 270\umsq{} per comparator
including shared rails (Fig.~\ref{fig:band}). Two mechanisms make the
band the unit of integration. First, its pins are carried to the
macro boundary on metal-3 risers with labels on the pin datatype, so
both the router and extraction bind them; earlier attempts to
integrate bare comparator cells failed precisely at pin binding.
Second, the band is small enough to flat-extract and LVS against its
own generated schematic, which is the check that caught two of the
three silent shorts of Section~\ref{sec:verification}. The signed-off
chip instantiates four bands (64 comparators, one per bitline of 32
columns); the TinyTapeout vehicle instantiates two.

\section{Verification Methodology}
\label{sec:verification}

The methodology section is deliberately a headline of this paper,
because the failures it catches are the ones that would otherwise
come back from the fab. The principle: \emph{every hand-assembled
structure is an independent claim about connectivity, and every claim
gets its own adversarial check.} Concretely, each level of the
assembly (bitcell, precharge row, sense band, array macro, full chip)
is flat-extracted with Magic and compared by netgen against a
transistor-level schematic generated from the same source data that
produced the layout. Builds are gated: a macro that does not reach
KLayout DRC zero \emph{and} netgen ``circuits match uniquely'' on its
own does not publish into the integration flow. On the compiler side,
the bit-exact chain of Section~\ref{sec:compiler} plays the same role
for numerics.

The discipline exists because chip-level signoff, even at DRC zero
and LVS zero, structurally cannot see certain defect classes inside
abstracted or merged geometry. Three real catches, summarized in
Table~\ref{tab:catches}, each DRC-legal and invisible at chip level:

\begin{table}[t]
\caption{Three silent shorts caught by per-level flat-extraction LVS.
All three are legal geometry to DRC; none is visible to chip-level
signoff.}
\label{tab:catches}
\centering
\footnotesize
\begin{tabular}{@{}L{2.35cm}L{2.75cm}L{2.35cm}@{}}
\toprule
Defect & Mechanism & Why signoff misses it \\
\midrule
Mask-jog abutment short & $+1$ via stack ends on the cell boundary,
abutting the neighbor column's $-1$ jog; even rows short BL$+_c$ to
BL$-_{c+1}$ & Magic merges touching same-layer tiles; shorts are
legal geometry to DRC \\
\addlinespace
N-well strip over latch NFETs & band-level well-merge strip overlaps
latch NFET diffusion; N$+$ in n-well extracts as a well tap, tying
internal nodes to VDD & band is abstracted at chip level; chip LVS
never sees band internals \\
\addlinespace
Tap tie on the HIT rail & a latch-up tap's local-interconnect bridge
lands on the comparators' HIT output li (an assumed power rail at
that height does not exist), tying all 16 outputs to VDD & legal
geometry; caught only by netlist compare of the band \\
\bottomrule
\end{tabular}
\end{table}

\textbf{Catch 1: mask-jog abutment.} The weight $+1$ programming via
stack originally ended exactly on the cell boundary, where it abutted
the neighboring column's weight $-1$ jog. Magic merges touching
tiles, so no DRC rule can ever flag the resulting connection; every
even row shorted its BL$+$ to the neighbor's BL$-$. Macro-level flat
LVS against the weight-derived schematic caught it; the mask
generator now keeps the stack west of the boundary with the $-1$ via
under a locally widened BL$-$ strip.

\textbf{Catch 2: the well strip.} The band builder's n-well merge
strip overlapped the latch NFETs' diffusion. N$+$ diffusion inside
n-well extracts as a well tap, so every comparator's internal nodes
were tied into the well/VDD network, in all prior builds. Chip
signoff had passed (the band is abstracted there); band-level
transistor LVS, added later, exposed it, and the fixed chip was
re-signed.

\textbf{Catch 3: the tap tie.} A band-level latch-up tap tie was
painted on local interconnect at a height assumed to hold a power
rail; the geometry there is actually the HIT output li, so all 16
comparator outputs were shorted to VDD: a dead sense band, fully
DRC-legal. Netlist compare caught it, but only after the defective
band survived one complete tile hardening because its LVS signature
(``no matching pin'' on every HIT) was misread as a known benign
label class. Two rules were institutionalized from this escape:
classify every LVS mismatch to a named net before dispositioning it,
and never paint band-level metal over sense-cell geometry without
tracing both shapes to their owning nets.

Three further flow-level catches belong to the same story. The
default timing flow modeled the analog VREF input with a logic
driving cell, so the resizer ``repaired'' the overloaded reference by
inserting a clock buffer: the comparators would have received a
digital rail instead of 0.9\,V (fixed with an analog-aware SDC and a
dont-touch on the net; the final netlist has zero gates on it).
Placeholder Liberty capacitances on the band starved the buffering of
real 130\,fF-class analog pins, fixed by generating the Liberty from
device-derived values. And the array's bitcell sources were found
floating (a deferred wiring item that behavioral checks could not
see) when the first full flat extraction ran, motivating the
source/strap network of Section~\ref{sec:macro}.

\textbf{DRC authority and calibration.} KLayout running the full PDK
deck (\code{sky130A\_mr.drc}, all options) is the authoritative DRC;
Magic's deck over-approximates several rule families by construction.
The divergence is calibrated rather than assumed: running the shuttle
precheck's exact Magic deck over already-fabricated shuttle projects
measures 544 items on a shipped analog design and 83{,}174 on a
shipped digital one, all in the wide-metal spacing family, while the
classes that gate manufacturability (latch-up, FEOL, tap spacing) are
zero on shipped work. That calibration, not optimism, is the basis on
which the Magic-only residue of Section~\ref{sec:results} is
dispositioned.

\textbf{Deterministic ECO closure.} The last few DRC items of a
routed chip are router artifacts (same-net notches, enclosed slivers)
whose textbook fix is a signoff ECO. Two mechanics make this
convergent and safe with open tools. Fills are generated
automatically from the DRC results database and are applied
\emph{inside} the flow's GDS stream-out step, never as a post-hoc GDS
patch: a whole-file round-trip re-streams macro internals, defeats
abstract extraction, and inflates LVS, whereas the in-flow patch
keeps extraction truthful (verified by byte-identical LVS reports
before and after fills). And because fills never perturb routing,
iterating fill generation against the stream-out tail converges in
one to two passes, where geometry-changing fixes would not.

\section{Results}
\label{sec:results}

\begin{table*}[t]
\caption{Signoff results of record (SKY130, open toolchain). The
chips share die, floorplan, flow, and read contract; the two
\code{cirom\_chip\_analog} runs differ only in weight mask; the
vehicles share the \texttt{test0} program. Chip STA at 25\,ns,
vehicle STA at 50\,ns; both chips read zero under the recorded-signoff
engine, and the parenthetical KLayout counts are current-engine
artifacts in unmodified foundry cells (see text).}
\label{tab:signoff}
\centering
\footnotesize
\setlength{\tabcolsep}{3.5pt}
\begin{tabular}{@{}lllll@{}}
\toprule
 & \code{cirom\_chip\_digital} & \code{cirom\_chip\_analog} ($\times$2 masks) & \code{tt\_um\_darga\_cirom} & \code{tt\_um\_azara\_cirom} \\
\midrule
Array & $64{\times}32$ ternary & $64{\times}32$ ternary & $64{\times}32$ ternary (test0) & $64{\times}32$ ternary (test0) \\
Readout & full-swing samplers & 64 comp. (4 bands) & samplers + on-die MAC & 32 comp. (2 bands) \\
Size & 0.85\,mm $\times$ 0.30\,mm & 0.85\,mm $\times$ 0.30\,mm & 161\um{} $\times$ 225.76\um{} & 161\um{} $\times$ 225.76\um{} \\
KLayout DRC & 0 (2 current-engine) & 0 (17 current-engine) & 0 (shipped GDS) & 0 (shipped GDS) \\
netgen LVS & 0 (``match uniquely'') & 0 (``match uniquely'') & 0 (``match uniquely'') & classified (see text) \\
Setup slack & $+16.3$\,ns & $+16.2$\,ns & $+29.4$\,ns & \\
Hold slack & $+0.11$\,ns & $+0.13$\,ns worst corner & $+0.11$\,ns & \\
Signal integrity & 0 slew/cap/fanout & 0 slew/cap/fanout & & \\
Functional & green, both masks & green, both masks & MVM + 64/64 raw reads & 64/64 rows reconstruct \\
\bottomrule
\end{tabular}
\end{table*}

\subsection{Full-Chip Signoff: Mask-Only Delta, Both Readout Styles}

\begin{figure*}[t]
\centering
\includegraphics[width=0.9\textwidth]{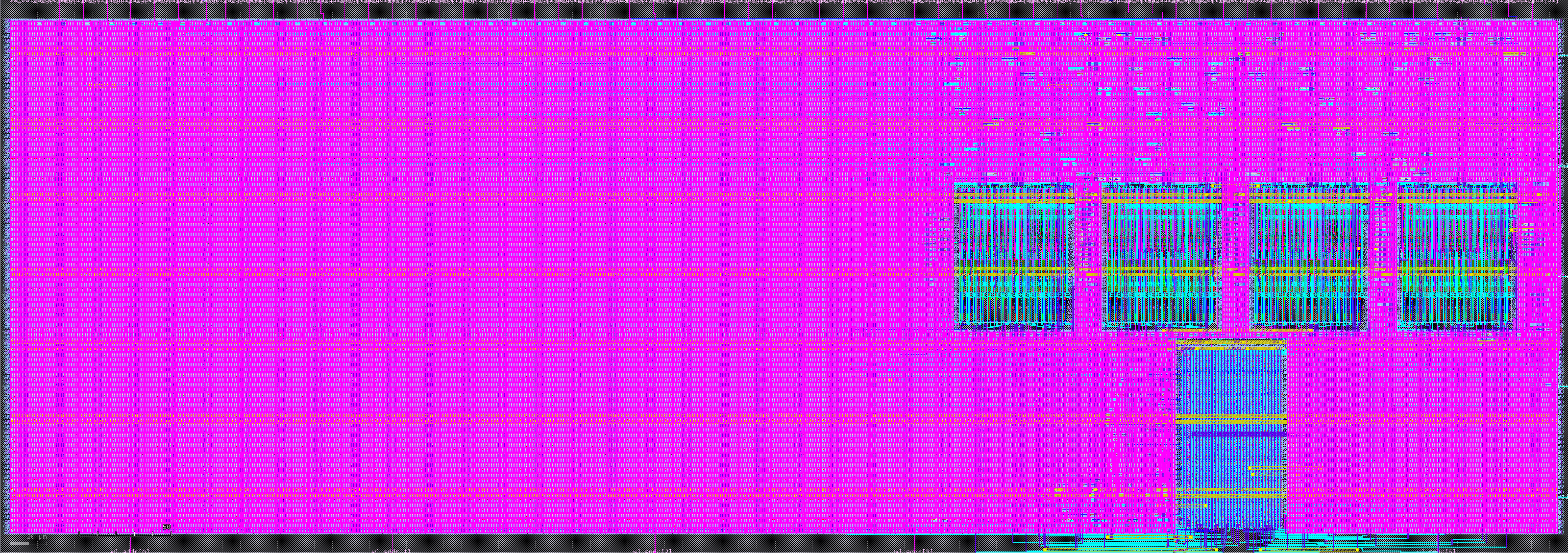}
\caption{The signed-off \code{cirom\_chip\_analog} (0.85\,mm $\times$
0.30\,mm, SKY130): $64\times32$ mask-programmed ternary array, four
16-comparator sense bands, read FSM. KLayout DRC 0, netgen LVS 0,
clean STA; signed off with two different weight masks through an
identical flow.}
\label{fig:chip-analog}
\end{figure*}

The signoff vehicle \code{cirom\_chip\_analog} (Fig.~\ref{fig:chip-analog})
integrates one $64 \times 32$ array macro, clocked precharge, 64
single-ended comparators in four bands, an external VREF pin, and a
read FSM, in 0.85\,mm $\times$ 0.30\,mm at a 25\,ns clock. It reaches
KLayout DRC zero, netgen LVS zero (``circuits match uniquely''), zero
max-slew/max-capacitance/max-fanout violations, $+16.2$\,ns of setup
slack, and $+0.13$\,ns of hold slack at the worst corner
(\code{min\_ff\_n40C\_1v95}; $+0.36$\,ns at nominal TT), with the
functional regression green (Table~\ref{tab:signoff}).

The load-bearing demonstration is that it did so \emph{twice}: once
with a checkerboard weight matrix, and once with a committed seeded
sparse matrix (\texttt{test0}: 500 cells of $+1$, 513 of $-1$, 1{,}035
of $0$) compiled through the same flow. Configuration, floorplan,
abstracts, and constraints are identical between the two runs; the
delta is the mask-programming geometry generated from the weight
file. Any $\{+,-,0\}$ matrix of the shape compiles to the same
signoff. The demonstration matrices are test patterns sized to the
hardened $64 \times 32$ shape; the frontend and bit-exact chain are
additionally validated on weight slices from the real 2B4T checkpoint
(Section~\ref{sec:compiler}).

\begin{figure*}[t]
\centering
\includegraphics[width=0.9\textwidth]{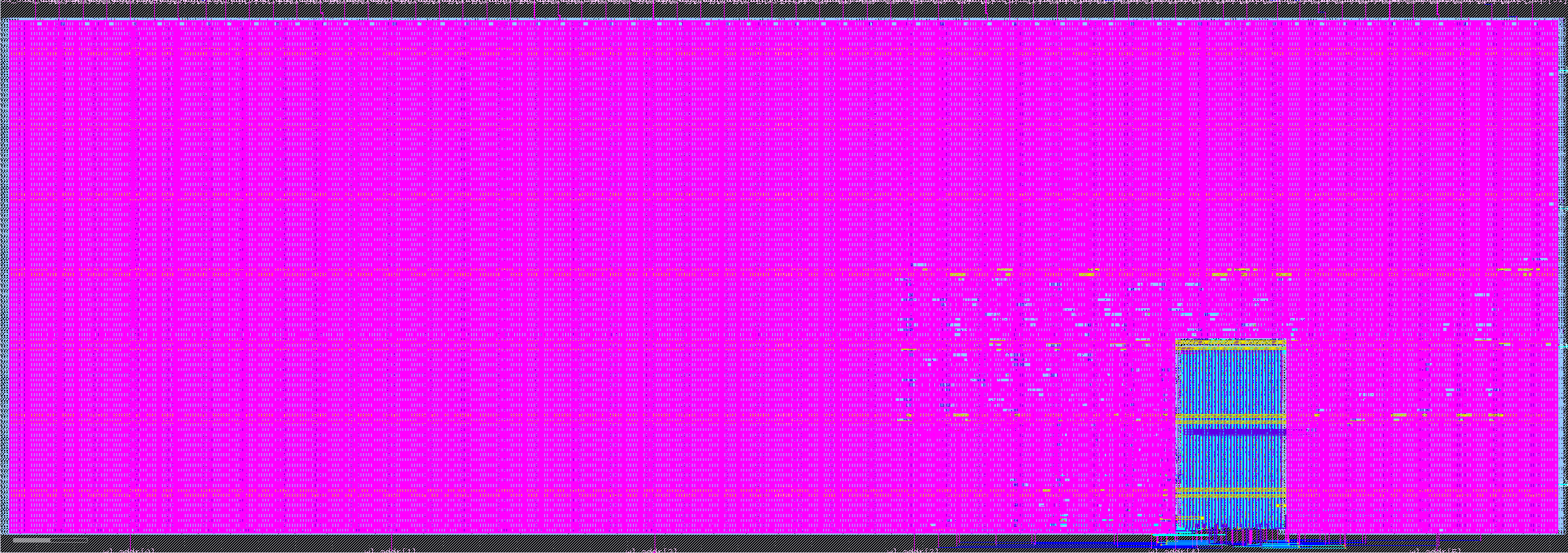}
\caption{The signed-off \code{cirom\_chip\_digital} (0.85\,mm
$\times$ 0.30\,mm, SKY130): the identical die, array placement, and
read contract as Fig.~\ref{fig:chip-analog}, with the four comparator
bands and the reference pin replaced by full-swing bitline samplers in
the standard-cell region.}
\label{fig:chip-digital}
\end{figure*}

The second axis of the demonstration is the readout style
(Fig.~\ref{fig:chip-digital}). A digital
variant of the same chip (\code{cirom\_chip\_digital}: identical
die, array placement, clock, and read contract, with the four
comparator bands and the reference pin replaced by full-swing bitline
samplers in standard cells) runs the identical flow to the same
gates: netgen LVS zero, $+16.3$\,ns setup and $+0.11$\,ns hold
slack, zero signal-integrity violations, and the functional
regression green on both weight matrices. Under the current rule
engine the digital chip reports two DRC items against the banded
chip's seventeen, all of them in a documented engine-artifact class
inside unmodified foundry cells; under the engine of the recorded
signoffs both read zero. The same die therefore signs off in both
readout styles, and the digital one measures cleaner.

\subsection{The Fabrication Vehicles}

\begin{figure}[t]
\centering
\includegraphics[width=0.62\linewidth]{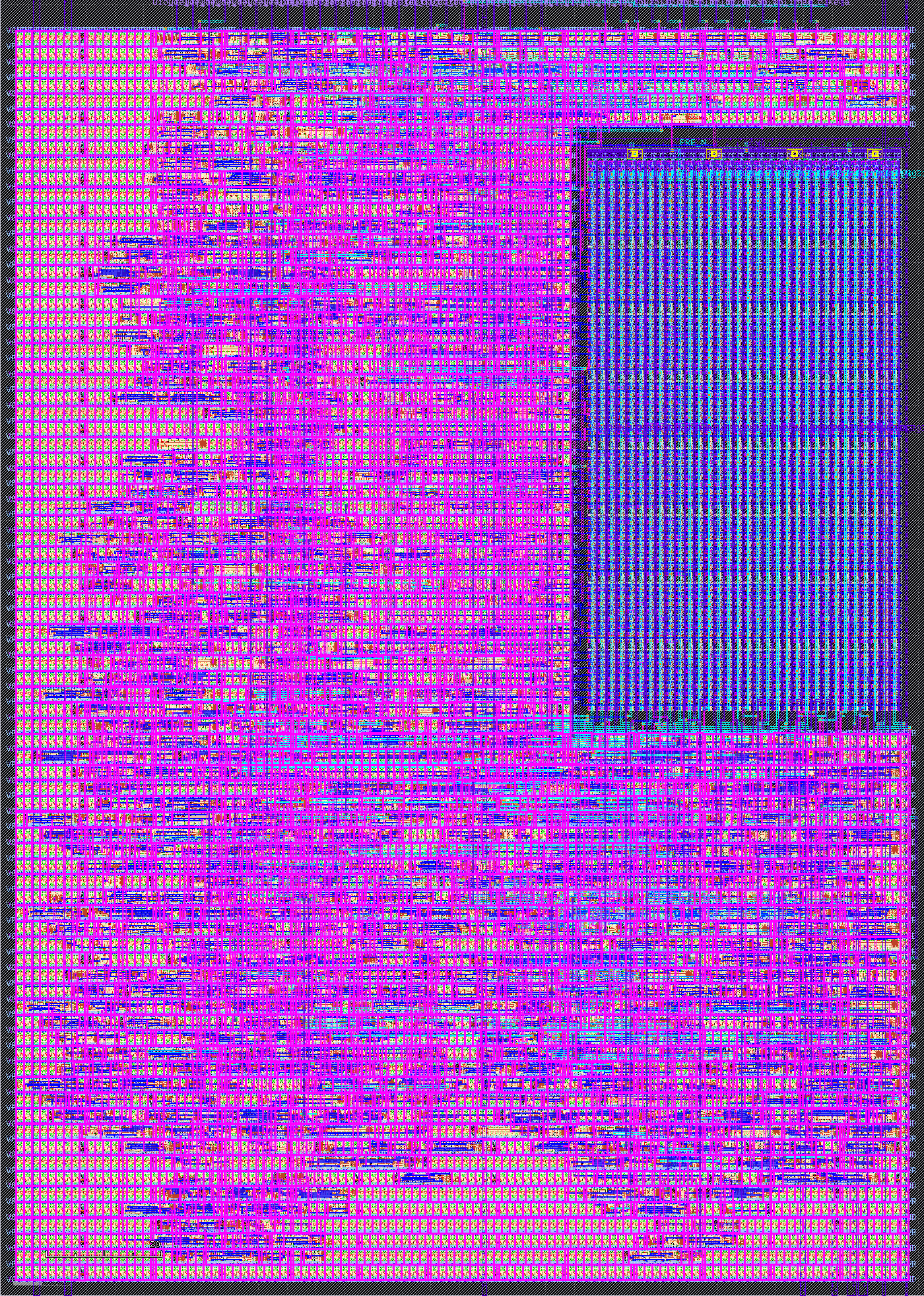}
\caption{The digital vehicle \code{tt\_um\_darga\_cirom} on a
TinyTapeout $1{\times}2$ tile (161\um{} $\times$ 225.76\um{}): the
mask-programmed array macro (top right, \code{test0} program),
with all 32 columns read by full-swing bitline samplers in the
standard-cell region in place of the two sense bands. No analog
pins.}
\label{fig:tt-digital}
\end{figure}

The digital vehicle \code{tt\_um\_darga\_cirom} is the
compiler's digital-tier emission hardened through the identical open
flow as a TinyTapeout-class tile
(Fig.~\ref{fig:tt-digital}): KLayout DRC zero under the full PDK deck,
netgen LVS matching uniquely, and static timing clean at all nine
corners, with 10{,}501\umsq{} of routed logic beside the same
5{,}345\umsq{} array macro. It carries the \emph{same} \code{test0}
mask program as the analog vehicle described next, making the pair a
controlled experiment: identical array, identical programming layer, readout
style as the only variable. The sensing-cost gap is also a coverage
gap made physical: the analog vehicle senses 16 of the array's 32
columns because two comparators per column do not fit, while the
digital tile senses all 32 because samplers do. The digital tile
additionally closes the loop the analog vehicle cannot, computing
ternary matrix-vector products on die (host-streamed activations; the
tile's vertical wiring budget does not carry an on-die activation
store, which the compiler retains for larger floorplans). The tile is hardened and its
submission repository passes TinyTapeout's hosted precheck.

\begin{figure}[t]
\centering
\includegraphics[width=0.62\linewidth]{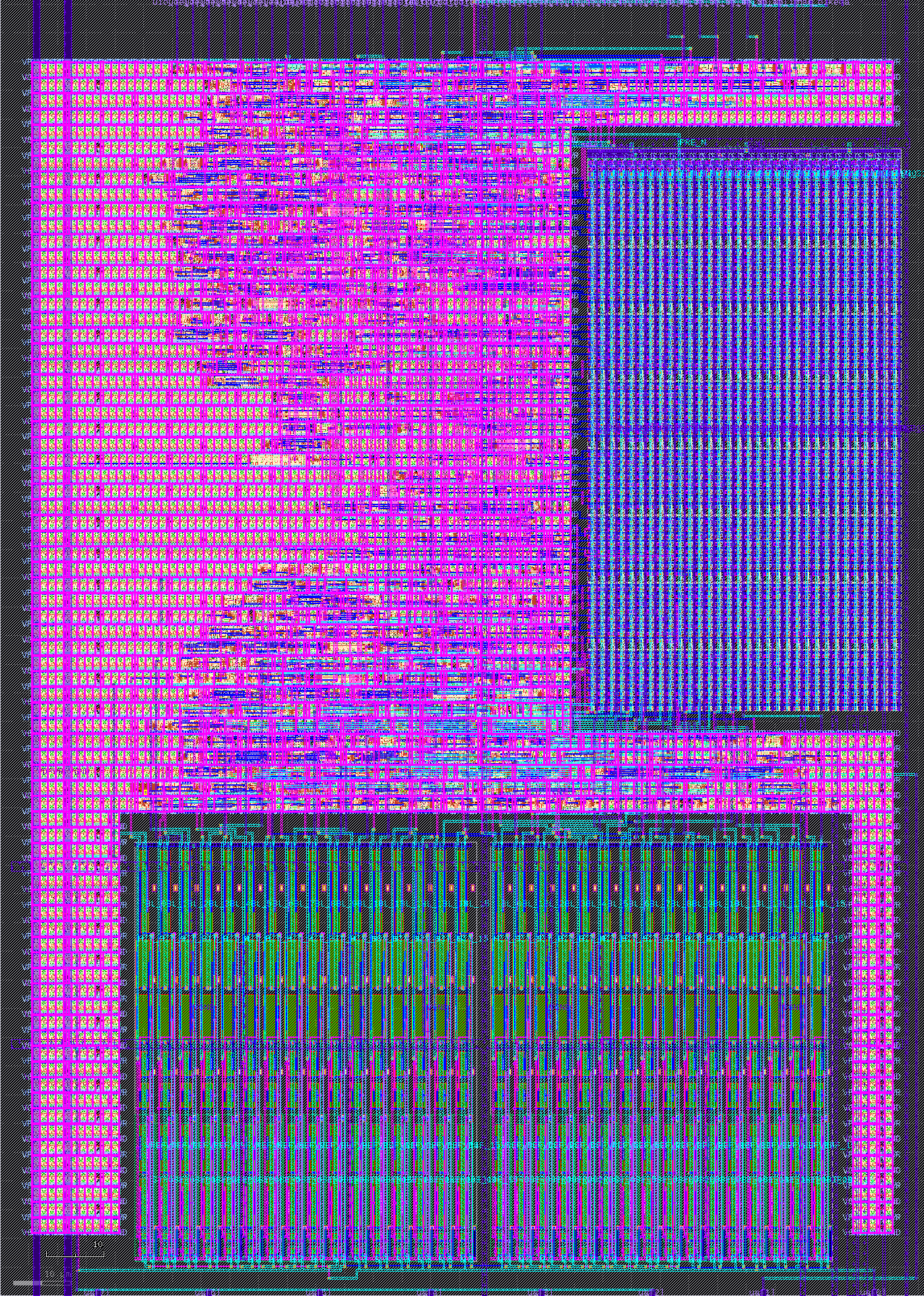}
\caption{The fabrication vehicle \code{tt\_um\_azara\_cirom} on a
TinyTapeout $1{\times}2$ tile (161\um{} $\times$ 225.76\um{}): array
(top), two sense bands, digital controller strip. External VREF on
\texttt{ua[0]}; column-0 bitline probes on
\texttt{ua[1]}/\texttt{ua[2]}.}
\label{fig:tt}
\end{figure}

The analog vehicle \code{tt\_um\_azara\_cirom}
(Fig.~\ref{fig:tt}) is a scaled subset of the signoff architecture on
a $1{\times}2$ analog tile (161\um{} $\times$ 225.76\um{}), targeting
the ttsky26c SKY130 shuttle~\cite{ttanalog}. It carries one $64
\times 32$ array with the \texttt{test0} matrix (2{,}048
mask-programmed weights). The program is chosen for testability, not
convenience: it is committed and seeded, so bench expectations
regenerate from the weight file; all 64 row signatures are distinct
(verified against the committed matrix), so address-decode faults and
pattern-systematic defects produce detectable mismatches that a
regular checkerboard would alias; and its composition (500/513/1{,}035
of $+1$/$-1$/$0$) is roughly half zeros, matching BitNet-class layer
sparsity so bitline loading is representative rather than a regular
best or worst case. On this vehicle, 16 of the 32 columns are sensed by 32
comparators in two bands in a single-strobe, fully parallel row read
(no analog multiplexing; serialization to the eight output pins is
digital and off the sense path). VREF is required externally on
analog pin \texttt{ua[0]}; \texttt{ua[1]}/\texttt{ua[2]} probe column
0's BL$+$/BL$-$, and the probe wiring adds an estimated 30--40\,fF to
that column's bitlines, making column 0 the deliberate worst case and
columns 1--15 the clean reference. A host drives a row address,
pulses start, and reconstructs the row's 16 ternary weights from four
output bytes; the bench reconstructs all 64 rows against the
weight-file expectations.

Submission status: routing DRC zero; the shipped GDS is KLayout-clean
under the full PDK deck; LVS is fully classified (both band instances
match uniquely inside the tile; the one residual class is the band's
p-substrate, exported as an anonymous port that lands on ground,
because the band intentionally carries no substrate taps and the
tile's tap network ties substrate to ground; verified benign on the
shipped GDS); and the shuttle precheck passes 14 of 15 checks, the
remaining one being Magic DRC whose count consists exclusively of the
wide-metal spacing family that fabricated shuttle projects carry at
scale (Section~\ref{sec:verification}). Latch-up and FEOL classes are
zero.

\subsection{Area Accounting}

Table~\ref{tab:area} gives the as-built area story. The honest
headline is that at 64 rows per bitline the sense periphery dominates:
two comparators per column cost about 540\umsq{} against about
160\umsq{} of column array, a $3.4\times$ tax. The tax is fixed per
column while the array grows with row count, reaching rough parity at
256 rows (Section~\ref{sec:scaling}).

\begin{table}[t]
\caption{As-built area accounting (SKY130). The per-column array cost
$\approx 2.5N$\umsq{} at $N$ rows follows from the 2.21\umsq{} cell
and the one-in-eight tap row.}
\label{tab:area}
\centering
\footnotesize
\setlength{\tabcolsep}{3.5pt}
\begin{tabular}{@{}lll@{}}
\toprule
Structure & Size & Normalized \\
\midrule
Bitcell & 1.70\um{} $\times$ 1.30\um{} & 2.21\umsq{}/weight \\
Array macro ($64{\times}32$) & 54.6\um{} $\times$ 97.9\um{} & incl. precharge \\
Sense band (16 comp.) & 59.3\um{} $\times$ 73.0\um{} & $\approx$270\umsq{}/comp. \\
Sense per column (2 comp.) & & $\approx$540\umsq{} \\
Array per column, $N{=}64$ & & $\approx$160\umsq{} ($3.4\times$ tax) \\
Array per column, $N{=}256$ & & $\approx$640\umsq{} ($\approx$0.85$\times$) \\
\bottomrule
\end{tabular}
\end{table}

\subsection{Estimator Methodology}
\label{sec:estimator}

Every cross-PDK number in this paper comes from one estimator chain,
so its structure and calibration are stated once. The area model is a
sum of five terms computed from the IR's layer dimensions and a
per-PDK descriptor: mask-programmed cells (weight count times an
anchored \umsq{}-per-weight density), periphery (per-block gate
counts anchored by synthesizing the actual controller and accumulator
RTL against the node's own libraries, the mechanism behind the
$206\times$ SKY130-to-ASAP7 anchor of Section~\ref{sec:asap7}),
KV-cache SRAM at the node's bit density, a fixed-function attention
engine, and a die-overhead fraction; each readout tier prices its own
sense periphery. The throughput model is a cycle count per token
(rows per pass at the configured bitline depth, column parallelism
from the tiled layout, pipelined steady state set by the slowest
layer, plus attention, KV-access, and inter-layer transport cycles)
under an effective clock. Reported configurations are the
minimum-area point of the tiling sweep, and all rates are
decode-fabric rates (Section~\ref{sec:asap7}).

The honesty of a number that cannot be measured lies in its bracket.
Three knobs dominate the cycle model, and each is back-fitted from an
independent evidence class rather than assumed: the clock derate
(achieved-over-ideal frequency as a function of area) from silicon
back-fits of open-PDK tapeouts plus placed-and-clocked anchors,
including this project's ASAP7 block and the independent HighTide
point~\cite{hightide2026}; inter-layer wire-transport cycles from
published interconnect scaling data~\cite{irds2024}; and KV-access
cycles from foundry SRAM datasheets and OpenRAM
measurements~\cite{openram}. Each fit emits low, mid, and high knob
values, every throughput below is reported as that bracket, and area
carries the stated bitcell-density sensitivity band. None of this
substitutes for measurement, which is why every such number is
labeled a calibrated estimate.

\subsection{Cross-PDK Model Estimates}

For scale context only, Table~\ref{tab:estimator} reproduces the
repository's calibrated estimator outputs
(Section~\ref{sec:estimator}); these are \emph{not} measurements.
The SKY130 rows use the
silicon-verified cell as their density anchor, the estimator models
the as-built 64-row bitline depth, and each row names its readout
tier (below 28\,nm only the digital tier exists, so the ASAP7 row is
digital). Three readings matter. First, chip area is
periphery-dominated at this node: the $6.8\times$ per-cell density
gap between the as-built cell and the BiROMA-bound anchor costs only
about 14\% of total area at the 22M-parameter reference shape.
Second, the readout tier is a first-order area term at model scale:
the digital tier estimates to 285\,mm$^2$ against the analog tier's
458\,mm$^2$ at the same shape and anchor. Third, no open shuttle
holds a real model at 130\,nm: the largest shape fitting an
815\,mm$^2$ die budget at the as-built anchor is about 23M
parameters. The value of the 130\,nm implementation is
reproducibility and validation of the contract, not deployment
density.

\begin{table}[t]
\caption{Calibrated model estimates (not measurements): a
22M-parameter BitNet-class shape, and the largest canonical model
shape fitting 815\,mm$^2$. The ``BiROMA-bound'' anchor is a smaller-node bracket
(Section~\ref{sec:biroma}).}
\label{tab:estimator}
\centering
\footnotesize
\begin{tabular}{@{}llrrr@{}}
\toprule
Anchor & Readout & Total & Cells & Periphery \\
\midrule
\multicolumn{5}{@{}l@{}}{\emph{22M-parameter shape (mm$^2$)}} \\
SKY130, as-built cell & analog & 458.2 & 48.4 & 334.3 \\
SKY130, as-built cell & digital & 285.2 & 48.4 & 183.9 \\
SKY130, BiROMA bound & analog & 401.2 & 7.1 & 326.1 \\
ASAP7 (anchored) & digital & 4.9 & 1.5 & 2.4 \\
\addlinespace
\multicolumn{5}{@{}l@{}}{\emph{Largest fit in 815\,mm$^2$ (params, mm$^2$), analog readout}} \\
SKY130, as-built cell & & \multicolumn{3}{r@{}}{22.9M at 530} \\
SKY130, BiROMA bound & & \multicolumn{3}{r@{}}{41.5M at 773} \\
\bottomrule
\end{tabular}
\end{table}

\textbf{Checkpoint coverage.} The frontend's claim of
model-agnosticism is exercised, not asserted: Table~\ref{tab:coverage}
runs published ternary releases from five organizations through the
same pipeline. Every checkpoint whose storage is exactly ternary
decodes through the strict ladder; the TriLM release additionally
carries the full product proof (compiled to a design bundle whose
6{,}912 mask programs audit bit-exact against the checkpoint and
whose RTL elaborates standalone), and the largest packed release
decodes completely at 7.0\,B ternary weights. One release family
stores \emph{group-scaled} ternary (each column group exactly
$\{-s_g, 0, +s_g\}$): the sign mask is recoverable, but a per-tensor
scale cannot carry the scale grid, so the frontend refuses it with
that diagnosis; faithful support is a per-group requantize in the
periphery, not a frontend change. Everything at 3.9\,B fabric
weights and below estimates inside one reticle-class ASAP7 die
($\sim$815\,mm$^2$); the $\sim$7\,B releases estimate just past one
reticle, the multi-die class.

\begin{table}[t]
\caption{Published ternary checkpoints through the pipeline
(storage detected per tensor; die area from the calibrated ASAP7
estimator: digital tier, mid bracket, 4096-token KV, 64-row bitline
depth, minimum-area tiling).}
\label{tab:coverage}
\centering
\footnotesize
\begin{tabular}{@{}lllr@{}}
\toprule
Checkpoint & Storage & Fabric weights & Die (mm$^2$) \\
\midrule
TriLM 190M & ternary fp16 & 0.113\,B & 57 \\
bitnet\_b1\_58-large & f32 QAT master & 0.679\,B & 210 \\
Falcon3-1B-1.58bit & packed u8 & 1.132\,B & 213 \\
Ternary-Bonsai-1.7B & group-scaled & 1.409\,B & 282 \\
Falcon-E-1B & packed u8 & 1.585\,B & 238 \\
BitNet b1.58 2B4T & packed u8 & 2.084\,B & 340 \\
Falcon-E-3B & packed u8 & 2.919\,B & 424 \\
bitnet\_b1\_58-3B & f32 QAT master & 3.222\,B & 671 \\
TriLM 3.9B & ternary fp16 & 3.681\,B & 759 \\
Falcon3-7B-1.58bit & packed u8 & 6.650\,B & 950 \\
Ternary-Bonsai-8B & group-scaled & 6.946\,B & 997 \\
Llama3-8B-1.58 & packed u8 & 6.979\,B & 988 \\
\bottomrule
\end{tabular}
\end{table}

\subsection{Simulated Read Energy, Power, and Throughput}
\label{sec:energy}

Dynamic read energy is the one quantity a capacitance-only extraction
is well conditioned to produce: it is $\sum CV^2$ over the switched
nodes, built from exactly the capacitances the extraction captures,
unlike delay, which needs the resistances we did not extract. We
therefore report simulated energy with more confidence than simulated
timing, and label both. The deck drives the C-extracted $64 \times
32$ macro (all 32 columns precharge and discharge according to the
mask) plus one \code{sa\_se} comparator pair at the FSM's 125\,ns
five-state cycle; per-comparator energy scales by the sensed-column
count, and every corner row is admitted only if both reads in the deck
decide correctly.

\begin{table}[t]
\caption{Simulated per-read energy on the C-extracted
checker-programmed macro (ngspice; corner, temperature, supply as
shown; ground return 500\,$\Omega$). Tile read $=$ macro $+ 16 \times$ SA pair. Leakage
is the idle precharge-hold state, macro plus one pair.}
\label{tab:energy}
\centering
\footnotesize
\begin{tabular}{@{}lrrrrr@{}}
\toprule
Corner & Macro & SA pair & Tile read & pJ/weight & Leak \\
       & (pJ)  & (pJ)    & (pJ)      &           & ($\mu$W) \\
\midrule
tt 27$^\circ$C 1.80\,V & 3.66 & 0.97 & 19.2 & 1.20 & 0.048 \\
ss 100$^\circ$C 1.62\,V & 3.07 & 0.79 & 15.7 & 0.98 & 0.052 \\
ff 125$^\circ$C 1.98\,V & 5.23 & 1.41 & 27.8 & 1.73 & 0.242 \\
\bottomrule
\end{tabular}
\end{table}

Table~\ref{tab:energy} summarizes. Three consistency checks anchor
the numbers. The supply energy arrives almost entirely in the recharge
phase (evaluate draws $<0.1$\,pJ because discharge dumps bitline
charge to ground; the rail pays when precharge restores it), which is
the correct physics for a dynamic read. The implied switched
capacitance, $3.66\,\mathrm{pJ}/V_{DD}^2 \approx 1.1$\,pF per read
across the columns, reproduces the $\sim$35\,fF-per-bitline figure
obtained independently from the extraction. And idle leakage is
properly corner-sensitive ($5\times$ from tt/27$^\circ$C to
ff/125$^\circ$C) once the measurement window sits past settling,
closing one deferred validation item: even at the hot-fast corner the
idle array holds its precharged bitlines at a $0.24\,\mu$W cost.

Rectified charge through the shared reference is
0.7--1.6\,pC per comparator pair per cycle: at continuous tile-rate
reads this is on the order of 150\,$\mu$A of average rectified
current through VREF, which is why both vehicles require a
low-impedance external reference and why the archived on-die divider
carries local decoupling in its design record.

Throughput needs no simulation: it is arithmetic on the verified FSM
and the signoff clock. The chip completes a read in five 25\,ns
states: 8.0M row reads/s, or 256M weight reads/s at full 32-column
sense. The tile adds byte serialization (ten cycles minimum per read,
five of them pin-budget overhead): 4.0M reads/s, 64M sensed-weight
reads/s at the same clock, with the analog front end itself sustaining
the chip's five-state rhythm. Combining the two: at continuous read
rate the tile's analog path averages $\sim$77\,$\mu$W, and a
vectorless OpenSTA report on the routed digital (activity 0.1,
tt/25\,ns, labeled a ceiling) adds 796\,$\mu$W, a sub-mW read
demonstrator. For scale only: Slim-Llama~\cite{kim2025slimllama} measures 9\,pJ per parameter
for a complete MAC datapath at 28\,nm; our 0.98--1.73\,pJ per weight
covers only the read primitive at 130\,nm, and ternary's digital
accumulation (AND-and-popcount, no multipliers) is the cheap part of
what remains. The estimator's token-throughput brackets stay withheld
here beyond noting they exist in the repository with their calibration
evidence.

\section{Two Readout Tiers: A Measured Comparison}
\label{sec:tiers}

Both tiers are the same machine. Accumulation is digital bit-serial
add/sub/skip in either case; the tiers differ only in how a
precharged bitline's state becomes a bit. The analog tier strobes a
clocked comparator against a shared VDD/2 reference
(Section~\ref{sec:sense}); the digital tier waits for the full swing
and samples the bitline into a flop through an input buffer. The
comparison below is on the same signed-off implementation, with each
number's provenance labeled.

\textbf{Area, measured.} The direct functional replacement for the
vehicle's two sense bands (8{,}656\umsq{}, hardened layout) is 32
bitline samplers, which synthesize to 1{,}081\umsq{} against the same
standard-cell library: $8.0\times$, or roughly 540 versus
68\umsq{} per weight column. The gap is the mismatch tax made
visible: the comparator's Pelgrom-sized input pair and its band
plumbing exist to resolve millivolts, and a full-swing sampler needs
none of it.

\textbf{Read energy, simulated (as built).} On the C-extracted deck
of Section~\ref{sec:energy}, the comparators are 81\% of the
19.2\,pJ tile read at the typical corner; the macro itself is
3.66\,pJ. Bitline energy is at parity by construction at this
geometry: hit bitlines discharge fully in either tier, and the
wordline holds through the strobe on the fabricated design, so the
analog tier realizes no partial-swing saving. As built, the digital
tier's total read energy is therefore roughly $4$--$5\times$ lower.

\textbf{The analog tier's remaining case, stated as a hypothesis.}
At production geometry (256-row bitlines, column-muxed sensing), a
comparator can in principle strobe at partial bitline depth and save
most of the swing energy that the full-swing sampler must spend. On
the fabricated design this saving does not exist (the wordline holds
through the strobe and evaluation is not terminated early), and
realizing it depends on three mechanisms no artifact implements:
evaluation terminated at the strobe, replica-timed strobing, and the
unfinished mux-resistance campaign. The hypothesis also reverses in
idle-dominated duty cycles until the reference is duty-cycled, since
the divider's static microamps outlast any per-read saving. The
March 2027 silicon decides: the analog vehicle returns a 64-point
measured offset map (the calibration anchor for the statistical sense
methodology) and measured read energy, against which the tier
selection is written down in the repository's roadmap with explicit
conditions. Until then the digital tier is the default emission and
the analog tier is a bounded, testable energy hypothesis carried by
its instrument.

\textbf{Why this order of evidence.} The analog tier was originally
selected on a necessity premise inherited from multi-bit
compute-in-memory, where the multiplier array dominates and analog
accumulation pays. Ternary invalidated that premise (the multiplier
is gone; accumulation was always digital here), and the premise was
retired only when the digital alternative was built and measured, not
argued. The comparison in this section is the paper's correction of
record, and the two-vehicle experiment is its test.

\section{A Negative Result: Two Ternary Weights per Transistor at
130\,nm}
\label{sec:biroma}

BitROM's BiROMA cell stores \emph{two} ternary weights per transistor
at 65\,nm~\cite{zhang2026bitrom}: each of the transistor's two
terminals connects by via choice to one of three per-side lines held
at $\{1/2, 1/4, 0\} \cdot \mathrm{VDD}$; zero is a connection (to the
mid line), never a float; in each read phase one side's lines are
driven while the other side's lines are merged and read as a voltage
level against comparators at $1/8$ and $3/8$ VDD. It is the single largest density
lever in the published ternary-CiROM space, so porting it to SKY130
was investigated seriously. The result is a documented no-go, in two
forks.

\textbf{Fork 1: discharge-domain, disproven topologically.} Keeping
our verified discharge sensing, the only pitch-competitive variant
shares one ``minus'' line per column between the two sides. This
fails structurally, not marginally. The storage transistor is
symmetric, so a cell stores the \emph{unordered} pair of its terminal
targets. With a shared minus line $SM$ and grounded zero strap $G$, the
configurations $(E{=}{-1}, O{=}0)$ and $(E{=}0, O{=}{-1})$ share the
target set $\{SM, G\}$: the identical circuit, so no read scheme
whatsoever can distinguish them (simulation confirms the two discharge a precharged $SM$
identically to within $5 \times 10^{-11}$\,V). It is worse than one
collision: $SM$ is shared down the column across both sides, and a
cell's opposite terminal is always low during a read, so any $-1$ on
either side discharges $SM$ in \emph{both} sides' sensing windows; a
shared minus line cannot attribute a discharge to the side being
sensed, leaving only $\{0, +1\}$ storable: fatal for ternary. Without
sharing, discharge sensing needs one line per readable nonzero value,
four sense lines per column, which re-inflates the pitch to
break-even ($\approx 1.03\times$ net density for the full periphery
and compiler rework).

\textbf{Fork 2: voltage-domain, feasible physics, uneconomic
periphery.} BitROM's actual scheme was then evaluated directly. The
core physics works at SKY130: a single NMOS delivers all three levels
through the read path, and the settled worst-case margin to the
nearest comparator threshold is 0.184\,V (slow corner,
100$^{\circ}$C, 63 off-cells loading the line), comfortably above
comparator offset (a $3\sigma$ of 30--45\,mV fits the 225\,mV band
without auto-zeroing). The blocker is the rails. The scheme requires
two low-impedance driven mid-rails ($1/4$ and $1/2$ VDD) that
\emph{source} the per-read settling charge; they are new analog
blocks (buffered rails plus decoupling) that the discharge design has
zero of, and their droop spends the same $1/8 \cdot \mathrm{VDD} =
225$\,mV band the margin lives in (our own high-impedance reference
measured 212--252\,mV of synchronous kickback: the entire band).
Compounding it, the slowest-settling level ($1/2$ VDD, weight zero)
is the most frequent symbol in zero-heavy BitNet matrices, and the
comparator count doubles. At 65\,nm BitROM reports its periphery at
4.8\% of macro area; at 1.8\,V 130\,nm the equivalent analog
periphery is a large fraction, refunding the $2\times$ cell density.

\textbf{Scope of the result.} This does not refute BitROM's 65\,nm
claim; it bounds the technique. The voltage-domain read is itself a
multi-level sense, so its viability window is set on both ends: on
the coarse end by 1.8\,V-class analog periphery costs (as measured
here), and on the fine end by the multi-level margin collapse of
Section~\ref{sec:scaling}. We estimate the window at roughly
65--28\,nm. The window bounds the \emph{analog} variant: the
two-weight gain requires resolving three levels on one line, so on
the fully digital readout it is unavailable at any node without a
multi-level digital discriminator, a different instance of the
architectural family. The dual-side mask-programming generator built for the
investigation (parameterized pitch, per-side via programming, DRC
zero and LVS-clean at spike scale) is retained for any smaller-node
port. The as-shipped design keeps one weight per cell.

\section{Scaling and Node Handoffs}
\label{sec:scaling}

\textbf{The sense tax and its amortization.} In the as-built design,
sensing is a fixed per-column cost: two comparators at
$\approx$270\umsq{} each, about 540\umsq{} per column, versus an
array cost of $\approx 2.5N$\umsq{} per column at $N$ rows
(Table~\ref{tab:area}). Fully parallel per-bitline sensing therefore
never amortizes at this node: $3.4\times$ overhead at $N{=}64$,
approximate parity at $N{=}256$, and longer bitlines stop paying
beyond that as the discharge window grows. The production path in the
repository's roadmap is 256-row bitlines with 8:1 column-muxed
single-ended \emph{binary} sensing, which models to a sense tax of
roughly 10--12\% (estimate, not silicon), explicitly gated on an
unfinished mux-resistance simulation campaign before it is trusted.
This mirrors mainstream ROM compiler practice of 4--16:1 column
muxing~\cite{openram} and the Taalas patent's statement that
simultaneous reads are limited only by sense amplifier count, that
is, muxing is by design~\cite{taalas2025patent}. The flop-based
digital sampler is the baseline this amortization must beat, measured
at roughly 68\umsq{} per column on the digital vehicle;
Section~\ref{sec:tiers} reports the as-built comparison and states
the 256-row analog case as the hypothesis it is.

\textbf{Three analog expiries.} The analog content of this design is
not one technique but three, and each has its own node horizon.
(1) \emph{Offset oversizing} (the 16\umsq{}-per-device input pair
behind the 270\umsq{} comparator) stops being the right answer at
roughly 28\,nm, where per-device $\sigma_{V_T}$ is smaller and an
auto-zero stage (about six switches and two capacitors for roughly an
order of magnitude of offset cancellation) is cheaper than area.
(2) \emph{Multi-level sensing} (BitROM's $1/8$ and $3/8$ VDD slicing,
and with it the BiROMA voltage-domain read) collapses at the
FinFET/low-supply boundary near 16\,nm, where $\mathrm{VDD}/8$ at
0.75\,V is roughly 94\,mV of absolute margin; BitROM demonstrates it
at 65\,nm and nothing published demonstrates it below.
(3) \emph{Charge-domain binary read} (precharge a line, let a cell
discharge it, strobe a clocked sense amp) survives to the end of the
roadmap: a 16\,nm production ROM senses this way~\cite{verma2015rom},
and the Taalas filing describes precisely this comparison at
N6~\cite{taalas2025patent}.

\textbf{The fully digital endpoint.} At 7\,nm-class nodes the
architecture that falls out is the Taalas-shaped one: mask ROM for
storage, a small clocked binary sense amplifier, and synthesized
digital everything else, including accumulation. Ternary moves this
crossover \emph{earlier} than for multibit CIM: a ternary MAC is an
AND gate and a popcount, with no multipliers, so the marginal benefit
of analog accumulation was always modest and the digital fabric
shrinks faster than the analog margin does. The compiler contract is
designed for this: the same IR, RTL, verification chain, and
estimators serve every node, while the contract's two custom cells
(bitcell, precharge) are re-laid-out per node, with the comparator
and replica-timing cells added only for the analog variant. A GF180MCU
bitcell timing characterization (discharge $1.25\times$ the SKY130
slow-corner value) exists as the first data point of that portability
argument.

\subsection{The ASAP7 Endpoint, Anchored}
\label{sec:asap7}

\begin{figure}[t]
\centering
\includegraphics[width=0.9\linewidth]{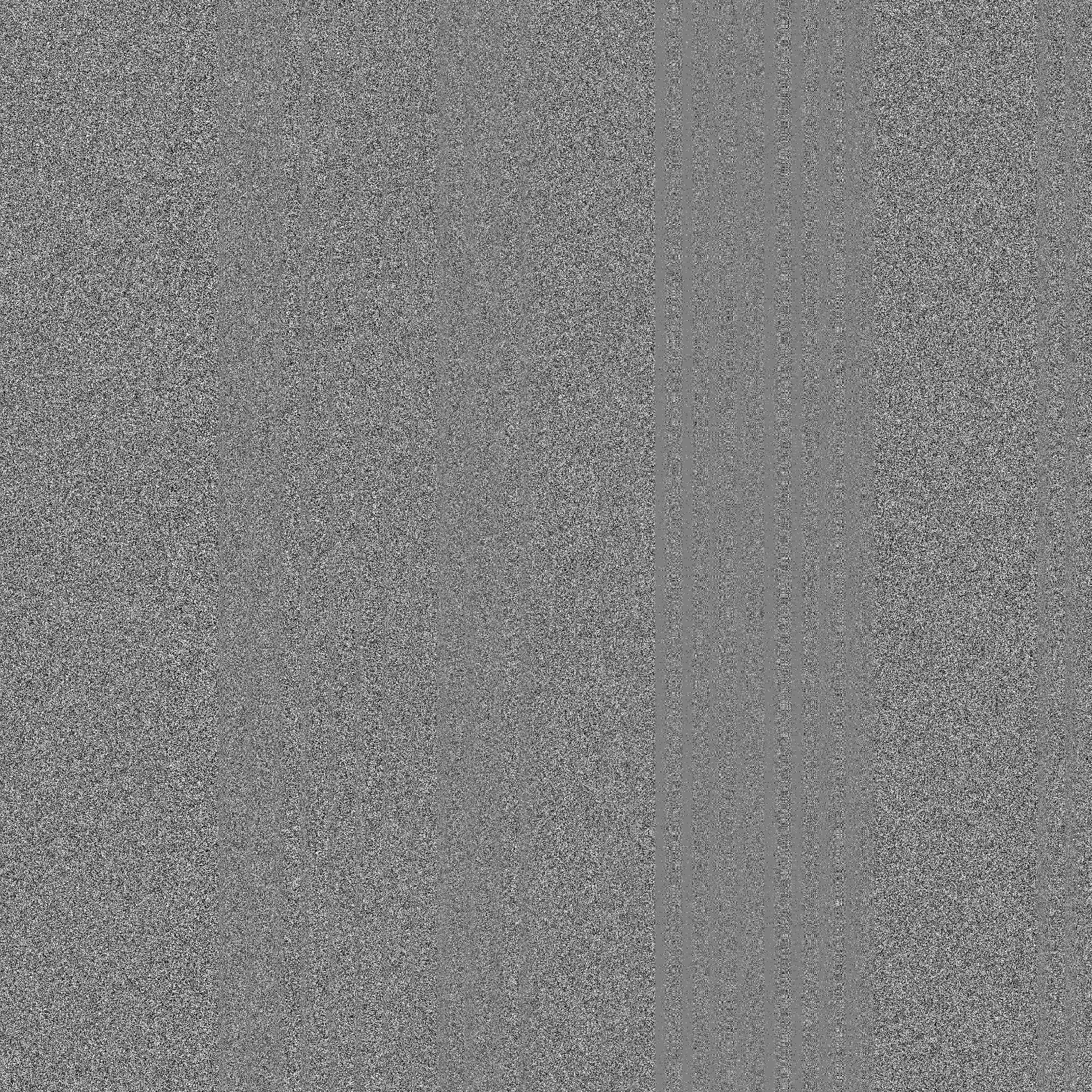}
\caption{One emitted layer's mask program (block-0 Q projection,
$2560 \times 2560$ weights, one pixel per weight) from the full-model
emission: 128{,}400 macros, 2.08B weights, verified bit-exact against
the checkpoint.}
\label{fig:emission}
\end{figure}

\begin{table}[t]
\caption{BitNet b1.58 2B4T bitline-depth sweep (calibrated ASAP7
estimates; A4 activations, 4096-token KV; 64-row is the
silicon-anchored geometry).}
\label{tab:depth}
\centering
\footnotesize
\begin{tabular}{@{}rrrr@{}}
\toprule
rows & mm$^2$ & tok/s & tok/s per mm$^2$ \\
\midrule
16 & 484 & 6.67M & 13{,}783 \\
32 & 390 & 4.21M & 10{,}801 \\
64 & 339 & 2.46M & 7{,}263 \\
128 & 312 & 1.33M & 4{,}282 \\
256 & 297 & 699k & 2{,}355 \\
512 & 289 & 358k & 1{,}237 \\
\bottomrule
\end{tabular}
\end{table}

ASAP7 is a predictive academic PDK, so these are physical-design
results on a community-standard 7\,nm proxy, not fabricable silicon;
the anchors bound the estimator, and the estimator carries the
claims. The ASAP7 estimator carries a measured synthesis anchor
(the digital-tier controller maps to 107.85\umsq{} against the ASAP7
RVT libraries versus 22{,}208\umsq{} at SKY130 through the identical
flow, $206\times$ for the same RTL), under which the estimator places
BitNet b1.58 2B4T's 2.08B-weight ternary fabric, with a 4096-token KV cache,
on 300--340\,mm$^2$ of a single reticle-class die at 0.36--2.5M
estimated tokens/s depending on bitline depth and activation width
(calibrated model estimates, not silicon). The 256-row/A8 point
decomposes as 145.9\,mm$^2$ of cells (2.08B weights at
0.07\umsq{}/weight), 68.1 of periphery, 50.3 of KV-cache SRAM, 1.7
of attention engine, and 32 of die overhead; the bitcell carries a
stated 0.05--0.10\umsq{} sensitivity band bounding the total at
roughly 250--370\,mm$^2$. At 7\,nm the bitline-depth tradeoff
inverts from the 130\,nm calculus: 64-row bitlines cost $+14\%$
area for $3.5\times$ the throughput, because the per-pass row window
rather than the sense periphery dominates, and the depth sweep has no
knee (area asymptotes at the cell-plus-KV floor while throughput
halves per depth doubling; Table~\ref{tab:depth}). Model risk
concentrates at the extremes:
deep points assume one row per 667\,ps clock regardless of bitline
RC, and shallow points imply macro counts the overhead fraction only
crudely captures. The rates are additionally decode-fabric rates,
not system rates: the estimator charges on-die periphery per token
(attention-engine, KV-access, and inter-layer transport cycles, each
a calibrated knob), but the off-fabric bf16 \code{lm\_head} and
sampling are the host's, and at the 64-row/A4 point the head alone
implies roughly 1.6\,PFLOP/s of sustained bf16 on that host. A
system claim would need the host budgeted; none is made here. Three artifacts anchor the
claim short of silicon: a 655k-weight CiROM block slice placed and
clocked through the ORFS ASAP7 flow (CTS+STA: 3.39\,GHz achieved,
13.9\,ps skew, 0.064\,mm$^2$; hold reads $-0.09$\,ns before the
standard repair pass and is reported as-is). The frequency class is
independently demonstrated on the node: the HighTide benchmark
suite's \code{partition\_o}, a different design by a different
group, closes at 3.41\,GHz post-route with the same tool stack, and
that point enters the estimator's calibration evidence
alongside our own~\cite{hightide2026}. The second artifact is
a full-model emission in which the
compiler tiles every ternary layer of the checkpoint into 128{,}400
mask-programmed macros (Fig.~\ref{fig:emission}; 2.08B weights, 0.93\% edge padding,
2.11\,GB of mask-program source, per-chunk digests) with the
bf16-tied \code{lm\_head} refused as off-fabric. The third is a
token-parity twin in which the transformers reference, a from-scratch
implementation, and the fabric's own readout-order decomposition
(bit-serial activation planes over 64-row groups, signed activations
offset-encoded onto the unsigned read) decode the model identically,
token for token. The digital shuttle vehicle
of Section~\ref{sec:tiers} carries the same contract to silicon
(Section~\ref{sec:tiers}), aimed at placing both emission tiers on
fabricated silicon.

\textbf{Weight custody and mask-level obfuscation.} The mask-only
invariant has a security corollary this paper does not pursue:
whoever holds the full mask set holds the checkpoint, so any
hardwired-model flow, at any node, inherits a custody problem. The
programming style here is unusually well placed to address it.
Because every cell carries an identical transistor and the weight
lives only in the via-1/metal-jog finishing geometry
(Figure~\ref{fig:bitcell-via}), the as-built array is already a
uniform fabric in which the model is resolvable only by imaging the
program layer itself: the mask-programmed counterpart of camouflaged
standard cells~\cite{rajendran2013camouflage}, obtained for free
rather than by dedicated camouflage libraries. The same confinement
makes split fabrication~\cite{imeson2013split} natural for CiROM:
fabricate every layer except the program metallization at one
facility and apply the program at another, and no single party
outside the compiling organization ever images the weights. Turning
the observation into a claim is future work: quantifying
reverse-engineering effort for via-programmed ternary arrays under
delayering, sharding the program across several masks so that no
single mask discloses the model, and the process-flow realities of
post-foundry program application. Nothing of the sort is demonstrated
on the present vehicle; the observation is only that CiROM's
programming style lets standard hardware-security techniques apply
directly to model weights, and that an open compiler is the natural
place to prototype them.

\section{Limitations}
\label{sec:limitations}

\textbf{No silicon measurements yet.} The fabrication vehicle is
submitted, not returned; silicon is expected in 2027, and
bench characterization (VREF sweeps of the comparator population,
column-0 discharge probing, sense-window mapping) is the subject of a
follow-up paper. Every silicon-adjacent number here is signoff,
extraction, or simulation, and is labeled as such.

\textbf{No measured energy.} Read energy and leakage are simulated
on a capacitance-only extraction (Section~\ref{sec:energy}) and the
digital power figure is a vectorless ceiling; nothing is measured,
resistive and short-circuit components are absent from the analog
numbers, and no efficiency comparison against measured silicon
(BitROM, TOM, Slim-Llama) is claimed.

\textbf{The analog vehicle demonstrates the read, not the MAC.} The
analog tile reads mask-programmed ternary rows through the sense
path; its accumulation chain (per-column popcount arithmetic,
bit-slice accumulate, between-layer requantize) is Verilator-verified
RTL exercised in the signoff chip's digital regression, not part of
that silicon. The hardened digital companion carries the ternary
matrix-vector product on die (host-streamed activations), but it is
submission-candidate, not fabricated; no on-die MAC is demonstrated
in silicon by either vehicle yet.

\textbf{One shape hardened.} The silicon-verified macro is the $64
\times 32$ shape; other shapes exist at the generator and
round-trip-flow level only, and no measurement at any deeper bitline
exists in either readout style, so the 128--256-row production
geometry is an extrapolation whose validation is explicitly gated. Sense margins rest on extracted-parasitic
simulation with capacitance-only extraction; hot-corner idle leakage
is now bounded by simulation (Table~\ref{tab:energy}), while mux
resistance for the 8:1 roadmap and replica-timed strobing remain open
validation items, and VREF is external on both vehicles.

\textbf{Scale honesty.} At 130\,nm density, real models do not fit
open shuttles (Table~\ref{tab:estimator}); the contribution is a
reproducible compiler-to-silicon path and its verification story, not
a deployable LLM chip.

\section{Conclusion}
\label{sec:conclusion}

Ankhdjet demonstrates that ``the model is a mask set'' can be an open,
reproducible artifact rather than a proprietary capability: a
HuggingFace ternary checkpoint compiles to a via-mask program of a
fixed, fully verified CiROM macro on SKY130, reaches KLayout DRC
zero, netgen LVS zero, and clean timing with open tools twice with
two different weight matrices, and a derived test vehicle is
submitted for fabrication on an open shuttle: to our knowledge the
first weights-to-mask CiROM compiler on a fabricable open PDK and the
first CiROM macro submitted for fabrication on one. The paper's
durable exports beyond the two firsts are methodological: per-level
adversarial LVS as the price of hand-assembled analog structures
(three DRC-invisible shorts caught), a measured readout-tier
comparison in which the digital sampler replaces the analog sense
function at one eighth the area and the same contract emits both
tiers (the digital companion hardened on the analog vehicle's own
mask program), a bounded negative result on two-weights-per-transistor
storage at 130\,nm, and a node map of where each analog sense
technique expires. Silicon returns in 2027 will test the sense-margin
story that extraction and Monte Carlo currently carry, and with it
the analog tier's remaining energy hypothesis.

\subsection*{Artifact Availability}

The compiler, cell generators, verification harnesses, flow
configurations, weight matrices, and signoff flows are open source
under the Apache-2.0 license at
\url{https://github.com/mpai17/ankhdjet}; the compiler and
calibrated estimators also install from PyPI
(\texttt{pip install ankhdjet}). Both chip signoffs and the shuttle
submissions regenerate from the repository's generators; the vehicle
submission repositories
(\url{https://github.com/mpai17/tt_um_darga_cirom},
\url{https://github.com/mpai17/tt_um_azara_cirom}) are generated
from it and carry TinyTapeout's hosted prechecks.

\section*{Acknowledgment}

This work builds on the open-silicon toolchain: Magic, KLayout,
netgen, ngspice, Yosys, OpenROAD, LibreLane, OpenRAM, the SkyWater
SKY130 PDK, and the TinyTapeout/ChipFoundry shuttle infrastructure.

\bibliographystyle{IEEEtran}
\bibliography{refs}

\end{document}